\documentclass[twocolumn]{aastex631}

\usepackage{CJK}
\usepackage{times}
\usepackage{amsmath}
\usepackage{graphicx}
\usepackage{hyperref}
\usepackage{gensymb}
\usepackage{upgreek}
\usepackage{natbib}
\usepackage{subfigure}
\usepackage{multirow}
\usepackage{comment}
\usepackage{mathtools}
\usepackage{mathrsfs}
\usepackage{fontenc}
\usepackage{color}
\usepackage{url}
\usepackage{pifont}

\newcommand{\Hb}{H$\beta$}
\newcommand{\MgII}{Mg\,\textsc{ii}}
\newcommand{\CIV}{C\,\textsc{iv}}

\newcommand{\sersic}{S\'{e}rsic}
\newcommand{\msersic}{$m_{\rm S\acute{e}rsic}$}

\submitjournal{ApJ}

\shorttitle{}
\shortauthors{Zhuang \& Shen}

\begin{document}
\begin{CJK*}{UTF8}{gbsn}

\title{Characterization of JWST NIRCam PSFs and Implications for AGN+Host Image Decomposition}

\author[0000-0001-5105-2837]{Ming-Yang Zhuang (庄明阳)}
\email{mingyang@illinois.edu}
\affil{Department of Astronomy, University of Illinois Urbana-Champaign, Urbana, IL 61801, USA}

\author[0000-0003-1659-7035]{Yue Shen}
\affiliation{Department of Astronomy, University of Illinois Urbana-Champaign, Urbana, IL 61801, USA}
\affiliation{National Center for Supercomputing Applications, University of Illinois Urbana-Champaign, Urbana, IL 61801, USA}

\begin{abstract}
We present a detailed analysis of the point spread function (PSF) of JWST NIRCam imaging in eight filters: F070W, F115W, F150W, F200W, F277W, F356W, F444W, and F480M, using publicly available data. Spatial variations in the PSF FWHM generally decrease with wavelength: the maximum and RMS fractional variations are $\sim20$\% and 5\% in F070W, reduced to $\sim3$\% and 0.6\% in F444W. We compare three commonly-used methods ({\tt SWarp}, {\tt photutils}, and {\tt PSFEx}) to construct model PSFs and conclude that {\tt PSFEx} delivers the best performance. Using simulated images of broad-line AGNs, we evaluate the impact of PSF mismatches on the recoverability of host galaxy properties. Host fluxes are generally overestimated when adopting mismatched PSF models, with larger overestimation for more AGN-dominated systems. Broader PSFs tend to produce less concentrated hosts while narrower PSFs tend to produce more concentrated and compact hosts. Systematic uncertainties in host measurements from PSF and model mismatches are generally larger than the formal fitting uncertainties for high signal-to-noise ratio data. Image decomposition can also lead to an artificial offset between the AGN and host centroids, which is common (e.g., $>1\sigma$ [$3\sigma$] detection in $\sim 80\%$ [$\sim 20-30\%$] of systems), and scales with the mean host surface brightness. Near the surface brightness limit, this artificial offset can reach as large as $\sim80$\%, 26\%, and 7\% of $R_e$ in systems with $R_e=$0\farcs12, 0\farcs48, and 1\farcs92, respectively. We demonstrate our PSF construction and image decomposition methods with an example broad-line quasar at $z=1.646$ in the CEERS field.

\end{abstract}


\section{Introduction} \label{sec1}

Tight correlations between the mass of supermassive black holes (BHs) and properties of their host galaxies, such as the stellar velocity dispersion \citep[e.g.,][]{Gebhardt+2000ApJ, Ferrarese&Merritt2000ApJ} and the luminosity/mass of the bulge \citep[e.g.,][]{Kormendy&Richstone1995ARA&A, Magorrian+1998AJ}, suggest that they may coevolve with each other \citep[e.g.,][]{Kormendy&Ho2013ARA&A}. The energy feedback from rapidly accreting active galactic nuclei (AGNs) is widely proposed to regulate the growth and evolution of itself and its host galaxy \citep[e.g.,][]{Somerville+2008MNRAS, Weinberger+2018MNRAS, Dave+2019MNRAS}. 

Due to the difficulties of measuring velocity dispersion from stellar absorption features, the studies of BH mass ($M_{\rm BH}$) and host property relations at high redshift mainly focus on the stellar mass ($M_*$) from broadband images or the dynamical mass as a surrogate from gas kinematics \citep[e.g.,][]{Peng+2006aApJ, Neeleman+2021ApJ}. In unobscured broad-line AGNs, $M_{\rm BH}$ can be estimated using emission from the broad-line region using the reverberation mapping technique or single-epoch methods \citep[e.g.,][]{Peterson+2004ApJ, Greene&Ho2005ApJ, Wang+2009ApJ, Shen+2015ApJS}. However, the emission from the accretion disk can easily dominate the total observed emission at rest-frame ultraviolet and optical wavelengths, hampering the measurements of host properties. Fortunately, image decomposition, which makes use of the different spatial extent of the AGN (unresolved) and its host galaxy (can be resolved depending on its size, redshift, and telescope), offers a feasible way to extract the emission from the host galaxy. 

At $z\gtrsim2$, even the reddest filter F160W of the Wide Field Camera 3 (WFC3) onboard the Hubble Space Telescope (HST) can only probe host emission at rest-frame $\lesssim5000$~\AA. At shorter wavelength, the AGN-to-host contrast is higher, rendering it more difficult to decompose host emission from the AGN. In addition, at rest-frame $\lesssim 5000$~\AA, host emission is dominated by the young stellar population and becomes insensitive to the total stellar mass. Meanwhile, emission from the host galaxy could be significantly blended with that of the AGN with $\sim$kpc ($\sim$0\farcs2) resolution. The unprecedented resolution (0\farcs03--0\farcs16), high sensitivity, and near-infrared (NIR) coverage up to $\sim5$ \micron\ of the Near Infrared Camera (NIRCam) onboard the James Webb Space Telescope \citep[JWST;][]{Gardner2006SSRv} make it possible to detect rest-frame $\gtrsim0.5-1.6$\micron\ emission of AGN host galaxies at $z>2$ for the first time. For example, with a full-width-at-half-maximum (FWHM) of the point-spread-function (PSF) of $\sim$0\farcs09 in the F277W filter, we can probe rest-frame $\sim0.93$ \micron\ emission from $z=2$ galaxies at a spatial resolution of $0.75$~kpc. The significantly improved resolution and rest-frame NIR coverage are essential to studying the properties of host galaxies of high-$z$ broad-line AGNs.

No instrument has the perfect PSF that does not vary with time and position on the focal plane, even for the NIRCam of JWST. Recent work has shown that the NIRCam PSF has strong spatial (up to $\sim15-20$\%) and temporal variations ($\sim3-4$\%) based on images of individual exposures \citep{Nardiello+2022MNRAS}. However, as AGN-host decomposition and most of the measurements are performed on images after combining individual exposures, it is important to analyze and quantify the properties of PSF for the final combined images. In this work, we compare the performance of commonly-used methods for constructing model PSFs and utilize the best PSF models to characterize the properties of the NIRCam PSF and their spatial variation. We then quantify the effect of PSF mismatch on the derived properties of AGN host galaxies from image decomposition using mock images of AGNs. As a demonstration of our overall methodology of PSF modeling and AGN-host decomposition, we present host galaxy measurements for a $z=1.646$ broad-line AGN. 

For this purpose, we use NIRCam imaging observations pointing in the south continuous viewing zone (S-CVZ; R.A.$=90.0000\degree$ and Decl$=-66.5607\degree$, i.e., the south ecliptic pole). This particular field has the advantage of having a much higher star density compared with other high-galactic-latitude fields, but much less source blending compared to observations of star clusters. These observations provide a large sample of stars from which we construct model PSFs and quantify their characteristics. 

This paper is structured as follows. Section~\ref{sec2} describes the data, data reduction, and PSF construction methods. Section~\ref{sec3} compares the performance of different PSF construction methods and quantify the properties of PSF and their spatial variations. Section~\ref{sec4} studies the effects of PSF mismatch on the recovered host properties. Section~\ref{sec5} applies our methodology to a real AGN. Our main conclusions are summarized in Section~\ref{sec6}. We adopt a cosmology with $H_0=70$ km s$^{-1}$ Mpc$^{-1}$, $\Omega_m=0.3$, and $\Omega_{\Lambda}=0.7$ and a \citet{Chabrier2003PASP} initial mass function (IMF).

\section{Data and Methods}\label{sec2}
\subsection{Data}

We use public NIRCam imaging observations from the Cycle 0 calibration program ``Stray Light Pointed Model Correlation'' (PI: Erin Smith, ID=1448) pointing at the S-CVZ field on 6th May 2022. The observations used Module B of NIRCam with the FULL subarray setting, covering a field-of-view (FoV) of $\sim$2\farcm2 $\times$ 2\farcm2. Module B consists of 4 short wavelength (SW; 0.6--2.3\micron) detectors and 1 long wavelength (LW; 2.4--5.0\micron) detector. SW detectors are separated by $\sim4-5^{\prime\prime}$, leaving gaps of cross shape at the center. Observations were taken in four filter pairs (SW+LW simultaneously) of three integrations each, with a total exposure time of 450.944~s for each filter pair: F070W+F277W, F115W+F356W, F150W+F444W, and F200W+F480M. A three-point dither pattern was used to optimize the subpixel sampling and to mitigate bad detector pixels. In particular, Mid-Infrared Instrument (MIRI) imaging was taken simultaneously with NIRCam imaging as coordinated parallel observation for this program. Therefore, a \texttt{3-POINT-WITH-MIRIF770W} dither pattern is adopted due to the requirement of a larger dither step by MIRI imaging.

\subsection{Data Reduction}

We use version 1.8.5 of the \texttt{jwst} pipeline with the Calibration Reference Data System (CRDS) version of 1039, with custom steps for data reduction. The detailed procedures are as follows: (1) We reduce the uncalibrated (\_uncal.fits files) raw data using the Stage 1 pipeline \texttt{calwebb\_detector1.Detector1Pipeline}, which applies basic detection-level corrections to each exposure, by adopting the default configuration with the exception of turning on the snowball correction. (2) We then apply $1/f$ noise (horizontal and vertical striping patterns) subtraction to the products of stage 1 pipeline (i.e., count rate data; \_rate.fits files) using scripts developed by the Cosmic Evolution Early Release Science Survey (CEERS; ERS 1345, PI: Steven Finkelstein) team \citep{Bagley+2022CEERS_data_reduction}. (3) We run Stage 2 pipeline \texttt{calwebb\_image2.Image2Pipeline} with default parameters, which involves steps including wcs assignment, flat-fielding, and photometric calibration and returns fully calibrated individual exposures (\_cal.fits files). (4) We perform 2-dimensional background subtraction on calibrated data after masking sources and bad pixels with a box size of 50 pixels using \texttt{SExtractorBackground} implemented in \texttt{photutils} \citep{photutils}. Astrometry is corrected using coordinates of sources from the GAIA data release 3 \citep[DR3;][]{GAIA_DR3} after accounting for their proper motions. 
(5) We combine individual exposures into mosaics using \texttt{resample} step in Stage 3 pipeline \texttt{calwebb\_image3.Image3Pipeline}. The output mosaics are drizzled to a common pixel scale of 0\farcs03 (30~mas) pixel$^{-1}$ without shrinking the input pixels (\texttt{pixfrac}$=1$) in all the filters following the choice by the CEERS team \citep{Bagley+2022CEERS_data_reduction}. All other pipeline parameters are set with their default values. The final round of background subtraction is performed to remove the remaining background (if any) in the mosaics. All the JWST data used in this paper can be found in MAST: \dataset[10.17909/6btv-br09]{http://dx.doi.org/10.17909/6btv-br09}.

\begin{figure}[t]
\centering
\includegraphics[width=0.5\textwidth]{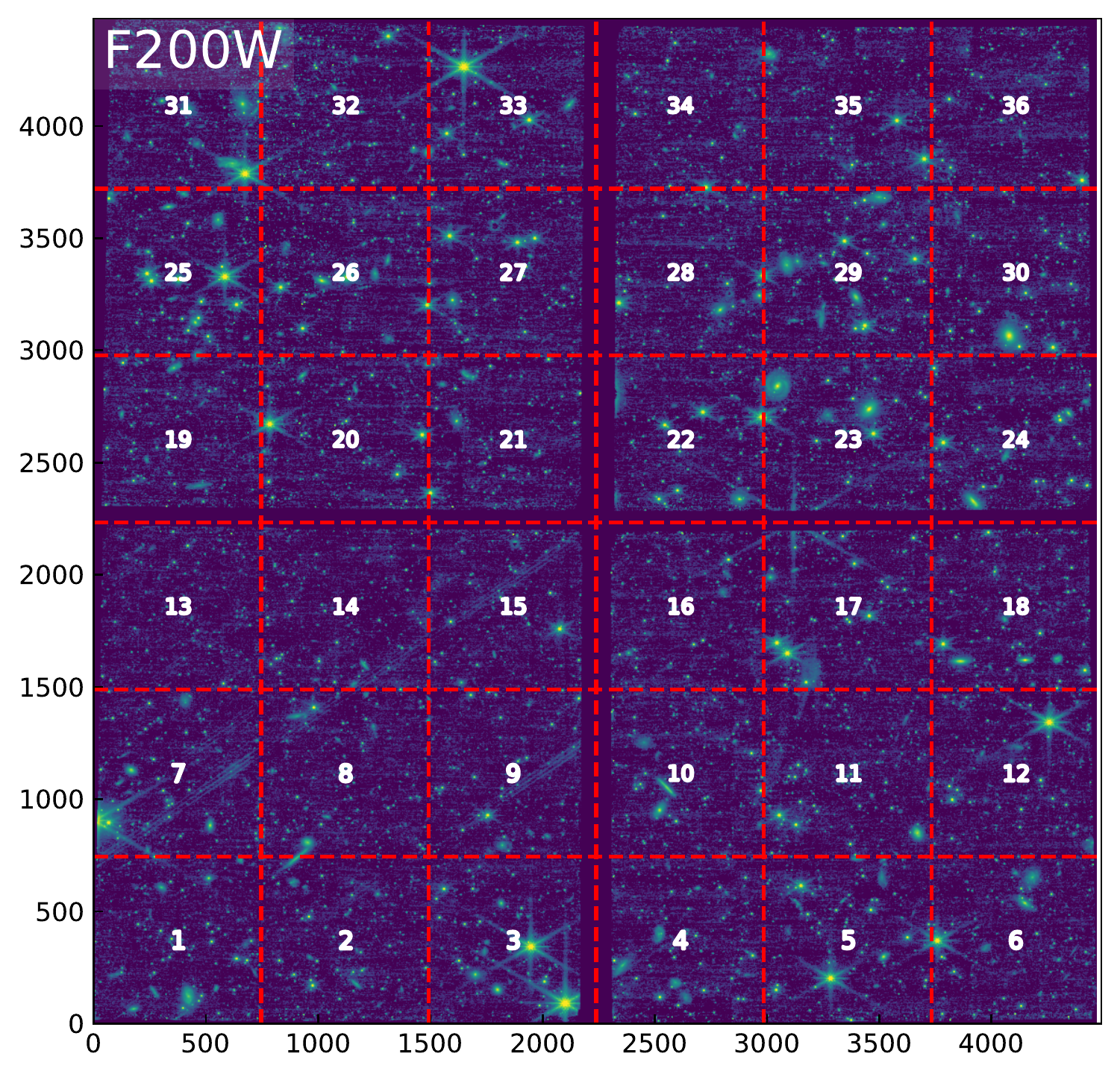}
\caption{Final F200W mosaic image with pixel scale of 0\farcs03 pixel$^{-1}$. Labels, vertical and horizontal dashed lines indicate the 36 regions for the characterization of the spatial variation of the PSF.}
\label{fig1}
\end{figure}

\subsection{PSF Construction}\label{sec2.3}

We use \texttt{SExtractor} \citep{1996A&AS..117..393B} to perform source detection and to select candidate PSF sources that are of high signal-to-noise ratio (SNR; \texttt{SNR\_WIN} $>100$), non-blended (\texttt{FLAGS} $<2$), non-irregular (\texttt{ELONGATION} $<1.5$), point-like (\texttt{CLASS\_STAR} $>0.8$), and without bad pixels (\texttt{IMAFLAGS\_ISO} $=0$). We then construct a master point source catalog by selecting objects that are detected in at least two adjacent filters. The final point source catalog consists of 1338 objects with \texttt{CLASS\_STAR} $\gtrsim0.96$ for 90\% of them. For each object in each filter, we generate cutout image with a size of 135 pixels ($\sim4^{\prime \prime}$) for SW images and 203 pixels ($\sim6^{\prime \prime}$) for LW images, which encircle at least 98\% and 97\% of the light \citep{Rigby+2023PASP}, respectively. Given the extremely high stellar density in this field (i.e., proximity to the Large Magellanic Cloud), the vast majority of these point sources are stars.

For high-resolution imaging analyses with HST or JWST data, it is customary to build empirical PSF models using a library of PSF sources. The major benefit of these PSF models is that they can mitigate the noise in the low surface brightness wings of the PSF, compared with PSF templates directly based on observed stars. We construct PSF models using two different methods: one by building a pixel-based model and the other by stacking PSF star images. Following \citet{Zhuang&Ho2022ApJ}, we use the Python package \texttt{photutils} \citep{photutils}, which is based on the algorithm of \citet{Anderson&King2000PASP}, and the software \texttt{PSFEx} \citep{Bertin2011ASPC} for the first method, and \texttt{SWarp} \citep{Bertin+2002ASPC} for the second method\footnote{we ignore the \texttt{reproject} package as its \texttt{reproject\_exact} function is currently known to have precision issues for images with resolutions $<0\farcs05$.}. We adopt default configurations for these packages, in particular a \texttt{quadratic} smoothing kernel for \texttt{photutils}, a \texttt{PIXEL} basis type for \texttt{PSFEx}, and a \texttt{lanczos3} resampling method for \texttt{SWarp}. 

We construct two types of PSF models for each method, one using all of the point sources across the entire FoV ({\em global}), and the other using point sources inside a particular region ({\em region}) for \texttt{photutils} and \texttt{SWarp} to probe variations of the PSF across the FoV. As illustrated in Figure~\ref{fig1}, we divide the final mosaic into 36 regions. One of the advantages of \texttt{PSFEx} is that it supports modeling of spatial variations of the PSF model across the FoV as a function of pixel coordinates ($x$ and $y$) in the form of a polynomial. In this work, we adopt a third-order polynomial with ten vector images ($constant$, $x$, $x^2$, $x^3$, $y$, $xy$, $x^2y$, $y^2$, $xy^2$, and $y^3$). For \texttt{PSFEx}, we construct the PSF model for each region using the central position of that region. 

The PSF models are constructed with an oversampling=2, {i.e., the pixel size of the PSF model is 1/2 of that of the input image (30~mas).} The final image size of PSF models is 265$\times$265 pixels for SW filters and 401$\times$401 pixels for LW filters. We verify that PSF models with oversampling=1 produce the same trends shown in Section~\ref{sec3}. The only exception is for \texttt{photutils}, oversampling=1 produces systematically larger FWHMs due to kernel smoothing, which occasionally fails, and in some cases produces unsatisfactory results.

\begin{deluxetable*}{cccccccccccccc}
\caption{Properties of PSF Models \label{table1}}
\tabletypesize{\small}
\tablehead{
\colhead{Method} & \colhead{Fitting size} & \colhead{FWHM} & \colhead{$q$} & \colhead{Nstar} & \colhead{FWHM [{\em region}]} & \colhead{$\sigma$(FWHM)}& \colhead{$q$ [{\em region}]} & \colhead{$\sigma$($q$)}\\
\nocolhead{} & \colhead{(pixel)} & \colhead{(mas)} & \nocolhead{} & \nocolhead{} & \colhead{(mas)} & \colhead{(mas)} & \nocolhead{} & \nocolhead{}\\
\colhead{(1)} & \colhead{(2)} & \colhead{(3)} & \colhead{(4)} & \colhead{(5)} & \colhead{(6)} & \colhead{(7)} & \colhead{(8)} & \colhead{(9)}
}
\startdata
\multicolumn{9}{c}{F070W}\\
\texttt{SWarp} & $ 7\times 7$ & 66.2 & 0.899 &  699 & (60.5, 67.5, 75.9) & 3.4 (5.0\%) & (0.818, 0.895, 0.974) & 0.031 (3.5\%)\\
\texttt{photutils} & $ 7\times 7$ & 69.6 & 0.914 &  700 & (64.3, 67.7, 77.4) & 2.9 (4.3\%) & (0.848, 0.911, 0.957) & 0.033 (3.6\%)\\
\texttt{PSFEx} & $ 7\times 7$ & 64.8 & 0.897 &  709 & (60.0, 64.5, 72.0) & 3.0 (4.7\%) & (0.829, 0.904, 0.985) & 0.041 (4.5\%)\\
\hline
\multicolumn{9}{c}{F115W}\\
\texttt{SWarp} & $ 7\times 7$ & 62.5 & 0.962 & 1020 & (58.2, 63.2, 68.3) & 2.1 (3.3\%) & (0.902, 0.957, 0.990) & 0.024 (2.5\%)\\
\texttt{photutils} & $ 7\times 7$ & 65.8 & 0.977 & 1020 & (61.6, 65.5, 68.8) & 1.5 (2.3\%) & (0.920, 0.968, 0.995) & 0.021 (2.2\%)\\
\texttt{PSFEx} & $ 7\times 7$ & 60.5 & 0.970 & 1018 & (58.0, 60.5, 64.0) & 1.6 (2.6\%) & (0.950, 0.967, 1.000) & 0.013 (1.3\%)\\
\hline
\multicolumn{9}{c}{F150W}\\
\texttt{SWarp} & $ 7\times 7$ & 65.2 & 0.966 & 1070 & (60.9, 66.5, 69.9) & 1.2 (1.8\%) & (0.928, 0.962, 0.988) & 0.015 (1.6\%)\\
\texttt{photutils} & $ 7\times 7$ & 69.2 & 0.980 & 1070 & (67.0, 69.1, 72.0) & 1.1 (1.6\%) & (0.930, 0.966, 0.990) & 0.013 (1.3\%)\\
\texttt{PSFEx} & $ 7\times 7$ & 64.7 & 0.969 & 1095 & (62.0, 65.0, 67.0) & 1.3 (2.0\%) & (0.954, 0.969, 1.000) & 0.008 (0.8\%)\\
\hline
\multicolumn{9}{c}{F200W}\\
\texttt{SWarp} & $ 7\times 7$ & 75.4 & 0.981 & 1058 & (72.9, 75.9, 78.0) & 1.0 (1.3\%) & (0.941, 0.980, 0.997) & 0.014 (1.4\%)\\
\texttt{photutils} & $ 7\times 7$ & 78.6 & 0.995 & 1058 & (76.3, 78.9, 80.9) & 1.1 (1.4\%) & (0.928, 0.972, 0.995) & 0.017 (1.7\%)\\
\texttt{PSFEx} & $ 7\times 7$ & 75.0 & 0.991 & 1080 & (74.0, 75.0, 77.0) & 0.9 (1.2\%) & (0.973, 0.987, 1.000) & 0.007 (0.7\%)\\
\hline
\multicolumn{9}{c}{F277W}\\
\texttt{SWarp} & $11\times11$ & 121  & 0.978 &  685 &    (113, 121, 129) &   3 (2.5\%) & (0.929, 0.973, 0.992) & 0.013 (1.3\%)\\
\texttt{photutils} & $11\times11$ & 122  & 0.983 &  685 &    (115, 121, 132) &   4 (3.3\%) & (0.893, 0.974, 0.997) & 0.030 (3.1\%)\\
\texttt{PSFEx} & $11\times11$ & 119  & 0.984 &  776 &    (115, 119, 122) &   2 (1.7\%) & (0.942, 0.975, 0.991) & 0.013 (1.3\%)\\
\hline
\multicolumn{9}{c}{F356W}\\
\texttt{SWarp} & $11\times11$ & 142  & 0.975 &  604 &    (136, 140, 151) &   2 (1.4\%) & (0.915, 0.977, 0.991) & 0.009 (0.9\%)\\
\texttt{photutils} & $11\times11$ & 140  & 0.982 &  604 &    (134, 140, 210) &   4 (2.9\%) & (0.655, 0.966, 0.987) & 0.019 (2.0\%)\\
\texttt{PSFex} & $11\times11$ & 138  & 0.983 &  583 &    (134, 138, 140) &   1 (0.7\%) & (0.957, 0.978, 0.993) & 0.008 (0.8\%)\\
\hline
\multicolumn{9}{c}{F444W}\\
\texttt{SWarp} & $13\times13$ & 162  & 0.993 &  323 &    (157, 161, 172) &   2 (1.2\%) & (0.950, 0.985, 0.994) & 0.010 (1.0\%)\\
\texttt{photutils} & $13\times13$ & 163  & 0.982 &  323 &    (159, 162, 171) &   3 (1.9\%) & (0.908, 0.969, 0.996) & 0.020 (2.1\%)\\
\texttt{PSFEx} & $13\times13$ & 160  & 0.991 &  323 &    (158, 161, 163) &   1 (0.6\%) & (0.963, 0.987, 1.000) & 0.010 (1.0\%)\\
\hline
\multicolumn{9}{c}{F480M}\\
\texttt{SWarp} & $13\times13$ & 178  & 0.980 &   52 &            \nodata &     \nodata &               \nodata &      \nodata\\
\texttt{photutils} & $13\times13$ & 180  & 0.984 &   52 &            \nodata &     \nodata &               \nodata &      \nodata\\
\texttt{PSFEx} & $13\times13$ & 178  & 0.980 &   48 &    (171, 179, 196) &   6 (3.4\%) & (0.905, 0.966, 0.994) & 0.016 (1.7\%)\\
\enddata
\tablecomments{Col. (1): Method used for PSF construction. Col. (2): Fitting region used for 2D gaussian fit. Cols. (3-4): Full-width-at-half-maximum and axis ratio between minor and major axis of global PSF model determined from 2D gaussian fit. Col. (5): Number of point like sources used for PSF construction. Col. (6) and (8): Minimum, median, and maximum values of PSF FWHM and q among 36 regions. Col. (7) and (9): $3\sigma$-clipped standard deviations of PSF FWHM and q among 36 regions, with fractional RMS in the parentheses (measured with respect to the median value across all 36 regions).}
\end{deluxetable*}

\begin{figure*}[t]
\centering
\includegraphics[width=0.49\textwidth]{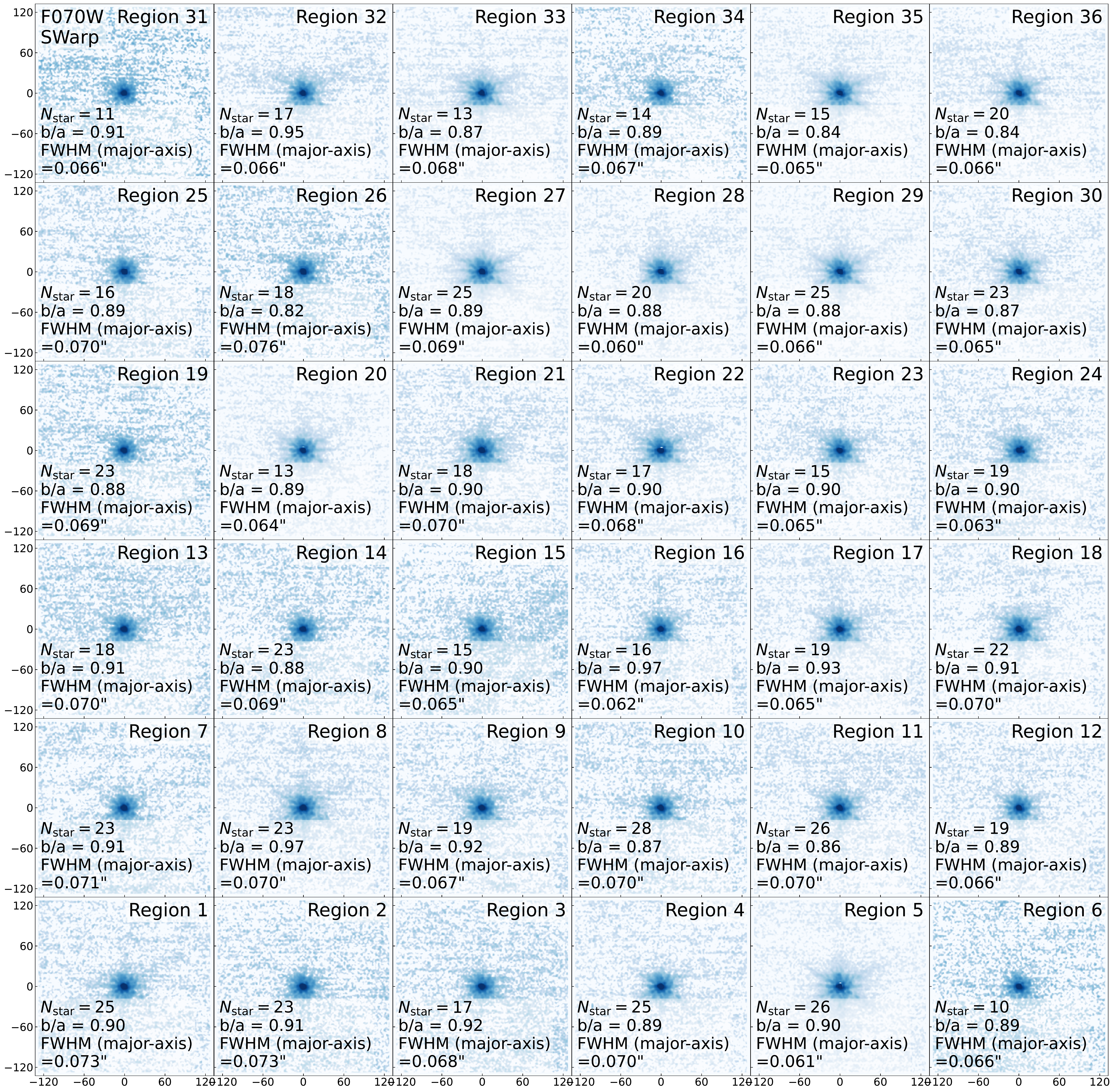}
\includegraphics[width=0.49\textwidth]{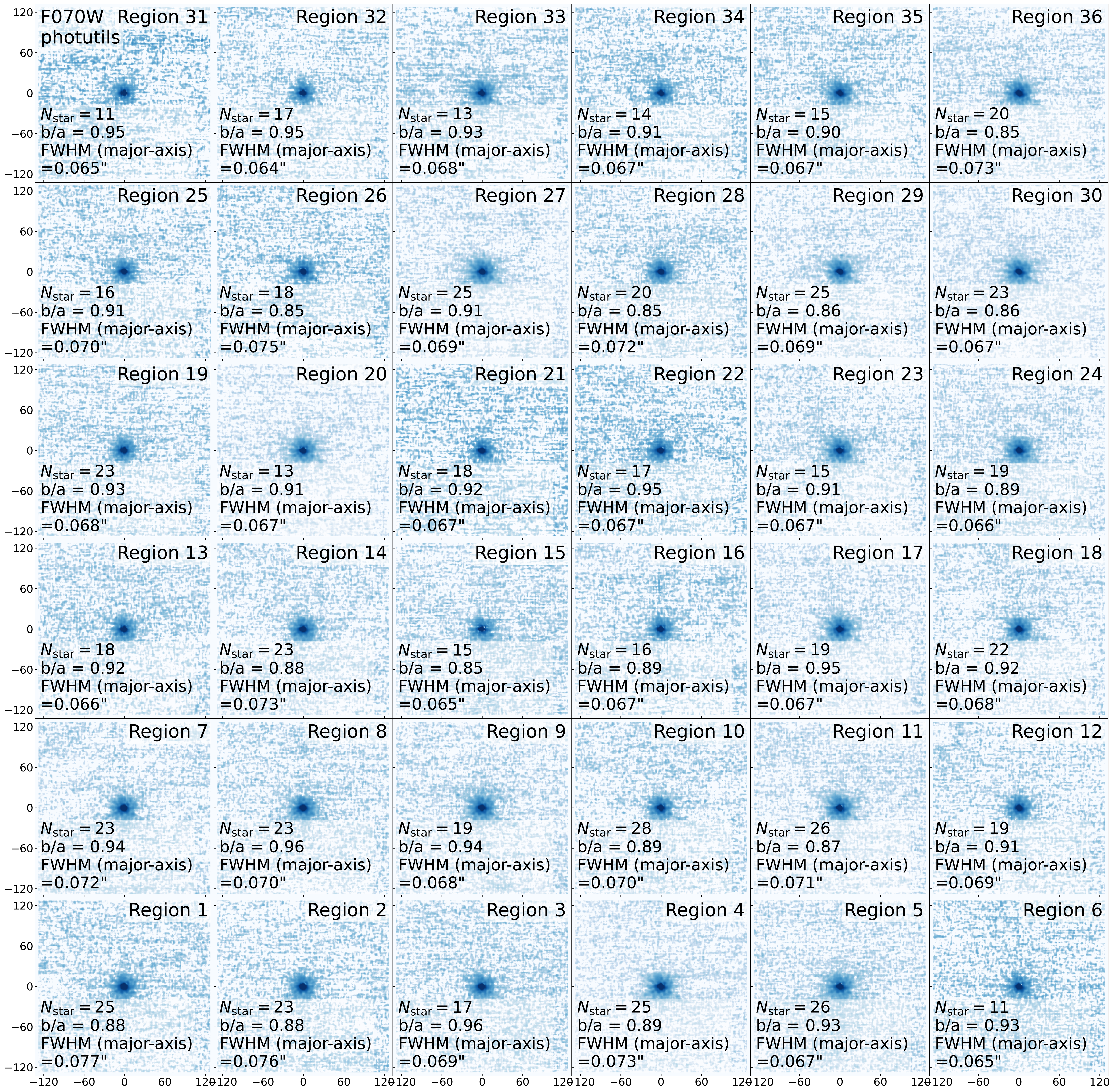}
\includegraphics[width=0.49\textwidth]{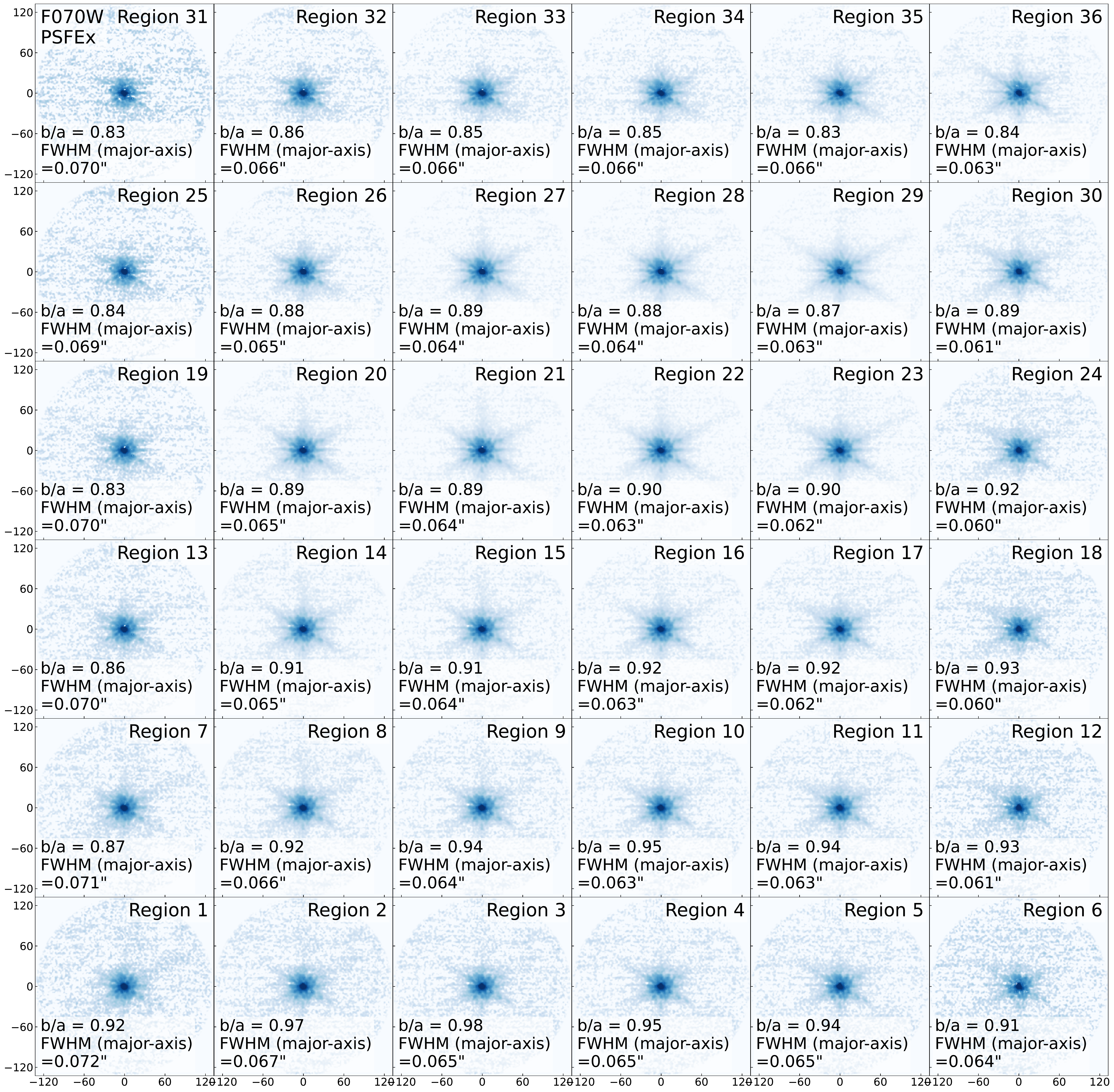}
\caption{The {\em region} PSF model images in different regions of the F070W filter constructed using \texttt{Swarp} (top-left), \texttt{photutils} (top-right), and \texttt{PSFEx} (bottom). The number of point sources used to construct the PSF model, axis ratio (minor-axis/major-axis), and FWHM along the major-axis are shown in the lower-left corner of each panel.}
\label{fig2}
\end{figure*}

\begin{figure*}[t]
\centering
\includegraphics[width=\textwidth]{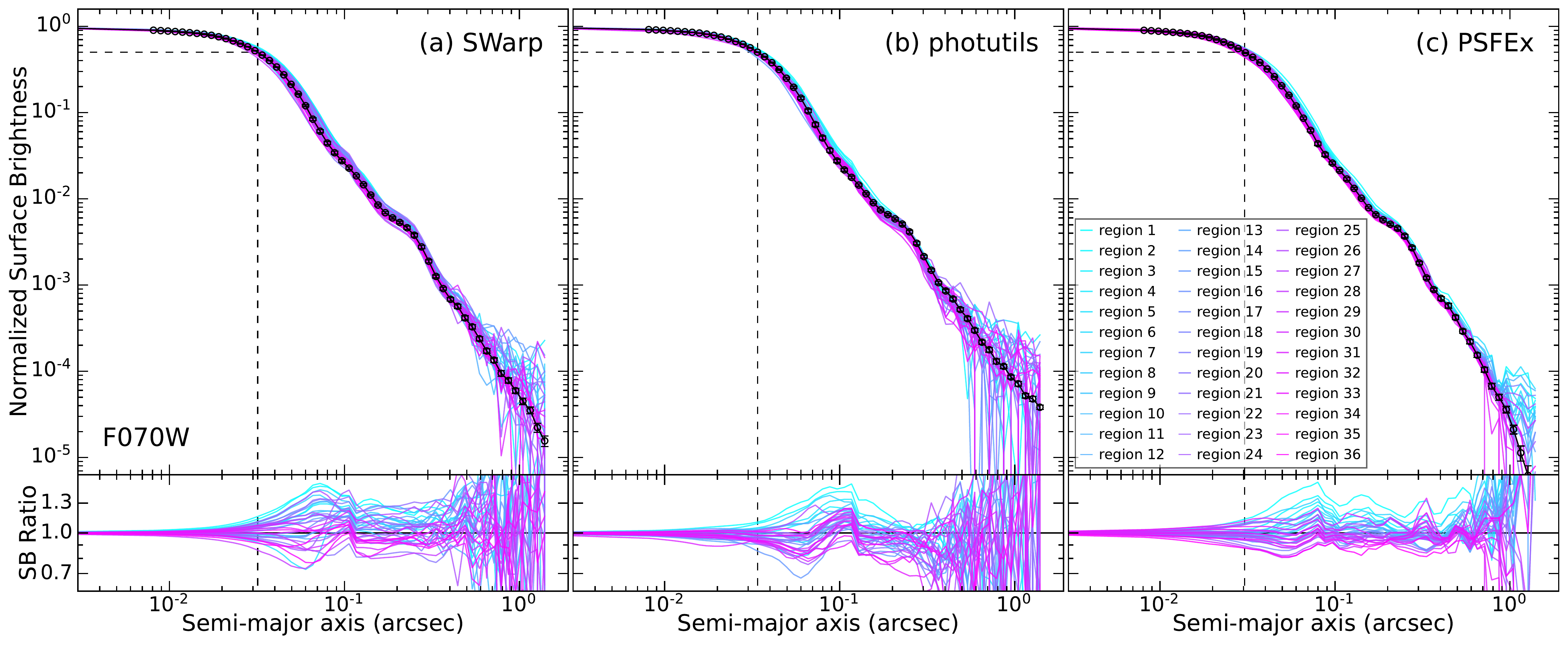}
\caption{Normalized radial surface brightness profiles of PSF models in the F070W filter at different regions generated using (a) \texttt{Swarp}, (b) \texttt{photutils}, and (c) \texttt{PSFEx}. Half-width-at-half-maximum of the {\em global} PSF is indicated with vertical black dashed lines.}
\label{fig3}
\end{figure*}

\begin{figure}[t]
\centering
\includegraphics[width=0.5\textwidth]{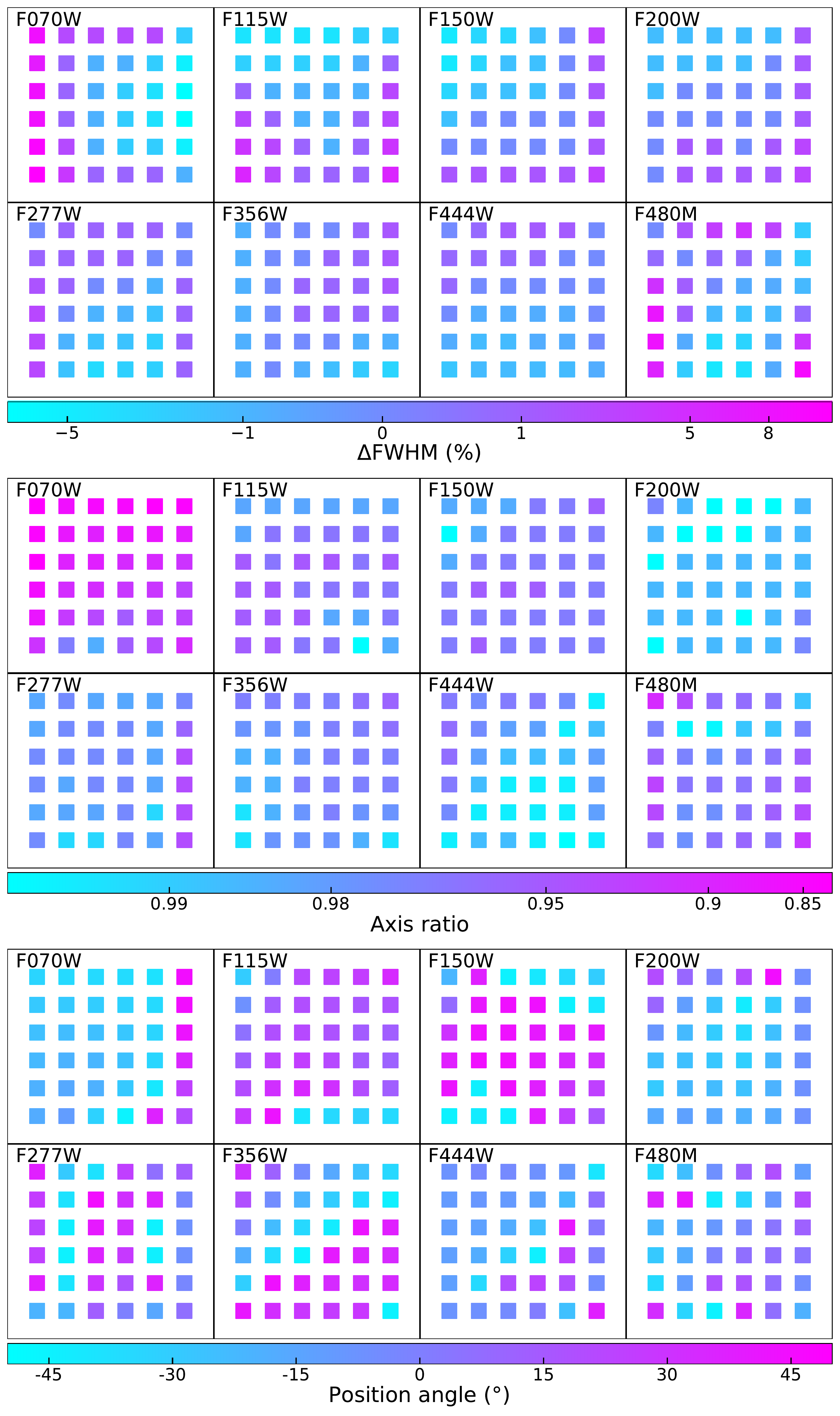}
\caption{Simultaneous AGN+host decomposition in six-band NIRCam images of the broad-line quasar SDSS1420+5300A at $z=1.646$. Images of data, model (AGN+host), data$-$nucleus, and residual (data$-$model) are shown from left to right. The best-fit parameters for the host galaxy (assuming a single Sersic profile for simplicity) are shown at the lower-left corner in the model panel. The right-most column shows the radial surface brightness profiles of data and models. The spiral arm features in the residual panel are faint and not included in the model.}
\label{fig4}
\end{figure}

\section{PSF Characteristics and Their Spatial Variation}\label{sec3}

\subsection{PSF Characteristics}\label{sec3.1}

We measure the shape of each PSF model by fitting a 2-dimensional Gaussian model to its core, with a fitting size of $2\times N +1$, where $N$ marks the index of pixel that is just larger than the half of the PSF FWHM. For {\em region} PSF models, we use the fitting size determined from the {\em global} ones. The geometric properties of the PSF models are presented in Table~\ref{table1}. Three methods generally produce consistent results on the geometry of the PSF models. PSF models constructed using \texttt{photutils} tend to have larger FWHMs in SW filters, likely due to the more significant kernel smoothing effect in less oversampled PSF models with small FWHMs ($\sim2$ pixels). 

Using the {\em global} PSF model as the typical representation for each filter, we find that PSFs in all SW and LW filters are critically- or well-sampled (FWHM$\gtrsim2$ pixels), given the drizzle parameters adopted in this paper (pixel scale=0\farcs03 pixel$^{-1}$ and \texttt{pixfrac=1}). Note that the FWHMs reported here are larger compared with those reported in the JWST user documentation\footnote{\url{https://jwst-docs.stsci.edu/jwst-near-infrared-camera/nircam-performance/nircam-point-spread-functions}}, especially toward shorter wavelength filters. As shown in \citet{Bagley+2022CEERS_data_reduction}, the recovered FWHM of PSF in the final mosaic is affected by the output pixel scale and \texttt{pixfrac}, with larger FWHMs from larger output pixel scale and larger \texttt{pixfrac}, due to increased pixel-to-pixel correlations. The PSF cores exhibit a high degree of axial-symmetry, with axis ratio (major-axis/minor-axis) $q\gtrsim0.97$ in all filters, except for the F070W filter ($q\approx0.9$).

We find clear spatial variations of the PSF across the FoV, as indicated by the spread of FWHMs of {\em region} PSF models from all three methods (Table~\ref{table1}). The degree of spatial variation depends on wavelength, with maximum fractional difference (maximum/minimum$-1$) and $3\sigma$-clipped standard deviation (RMS) of the PSF FWHM decreasing from $\sim20$\% and 5\% in the F070W filter to $\sim3$\% and 0.6\% in the F444W filter. This level of variations is similar to that reported by \citet{Nardiello+2022MNRAS}. 

Using F070W as an example, {Figure~\ref{fig2} and Figure~\ref{fig3} show the {\em region} PSF model images across the FoV and their normalized radial surface brightness profiles, respectively}. There are significant radial variations up to 100\% at $60-90$ mas ($2-3 \times$ FWHM) from the center. The variations persist toward larger radii and are dominated by random noise at $r\gtrsim1$\arcsec\ due to background fluctuations. Thanks to the simultaneous modeling of point sources across the FoV, PSF models constructed using \texttt{PSFEx} have much higher SNR at large radii compared with the other two methods, with less contamination from background noise and low-level, small-scale gradients that are not accounted for by background subtraction. Together with better performance of PSF models constructed from \texttt{PSFEx} in terms of modeling the light profiles of point sources, as well as measuring their magnitudes (Section~\ref{sec3.2}), we use PSF models constructed using the \texttt{PSFEx} method as the reference in the following analyses.

Figure~\ref{fig4} shows the spatial variations of FWHM, axis ratio, and position angle of the PSF models in eight filters across the FoV of NIRCam. We find that the PSF models with the broadest and narrowest profiles are preferentially located toward the edge of the FoV. Different filters show distinct patterns: a horizontal pattern in F070W, a vertical pattern in F115W, F277W, F444W, and F480M, and a diagonal pattern in F150W, F200W, and F356W. These patterns could be affected by stochastic aliasing of the PSF during the process of drizzling, which is more significant for SW filters where the output pixel scale is close to the input value with experience from HST \citep{Rhodes2007ApJS}.
Axis ratio and position angle exhibit somewhat less spatial variation, with extreme values more commonly found toward the corners. The variation in position angle is partially due to the large uncertainty of the position angle measurement at large axis ratios.

It is possible that the spatial variation of the PSF shown in this work is somewhat underestimated due to the smoothing from the adoption of the third order polynomial function for the \texttt{PSFEx} (Section~\ref{sec2.3}). {While these spatial variations of the NIRCam PSF are modest in most filters (e.g., fractional RMS of a few percent), they may impact the detection and measurements of host galaxies in unobscured AGNs where the light is dominated by the centrarl AGN. Quantifying the effects of PSF mismatches on AGN+host image decomposition is one of the main motivations of this work (Section~\ref{sec4}).} 

\begin{figure*}[t]
\centering
\includegraphics[width=0.8\textwidth]{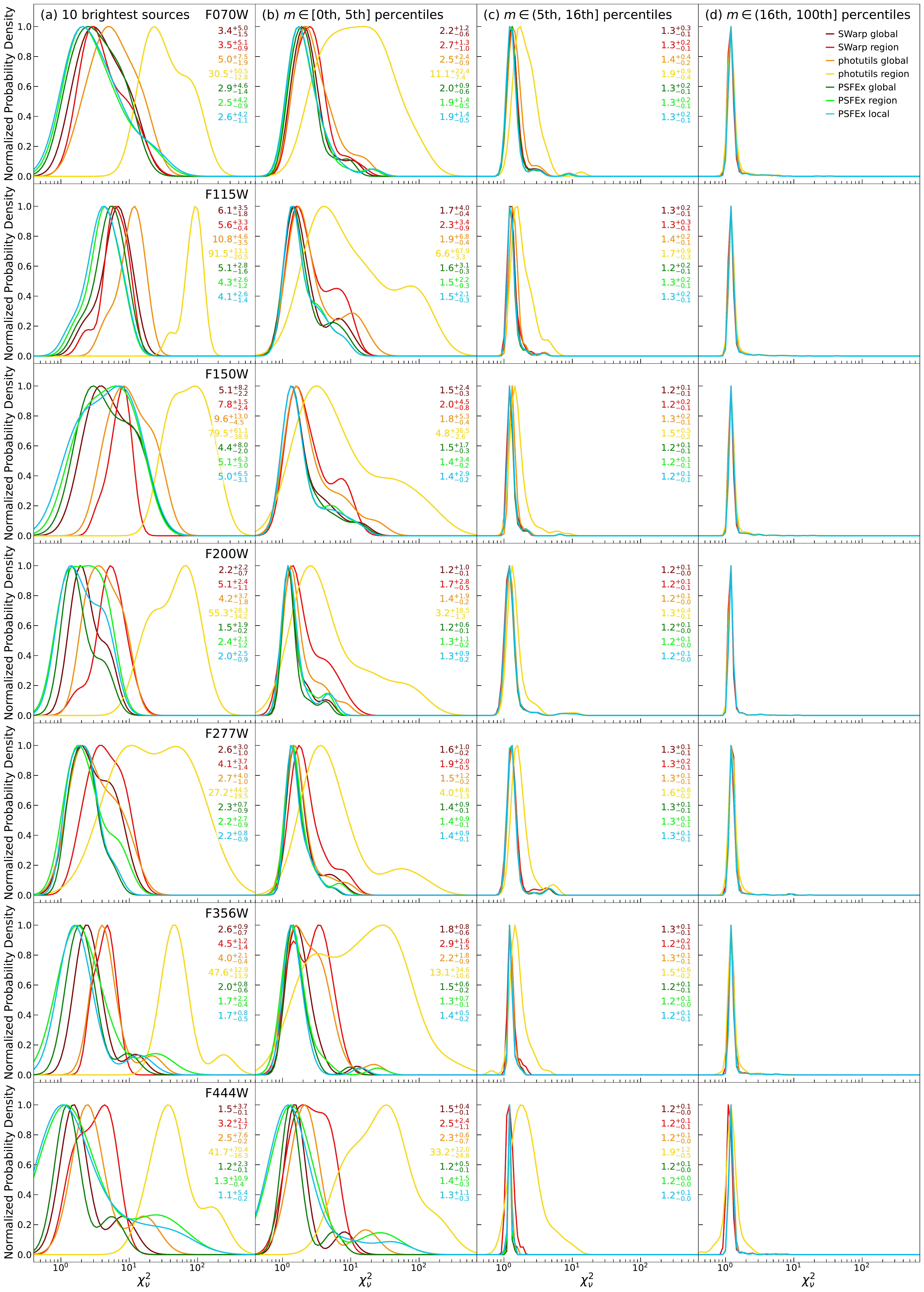}
\caption{Normalized probability density distributions of reduced $\chi^2$ ($\chi^2_{\nu}$) of fitting to point sources using PSF models constructed from different methods, estimated from Gaussian kernels using \texttt{scipy.stats.gaussian\_kde}. Maroon, red, orange, gold, green, lime, and blue lines represent results from \texttt{SWarp global}, \texttt{SWarp region}, \texttt{photutils global}, \texttt{photutils region}, \texttt{PSFEx global}, \texttt{PSFEx global}, and \texttt{PSFEx local}, respectively. \texttt{PSFEx local} represents the PSF models constructed at the exact location of the source in the image. Columns (a--d) indicate results for (a) 10 brightest sources, sources with magnitude between (b) 0th and 5th percentiles, (c) 5th and 16th percentiles, and (d) 16th to 100th percentiles, respectively. Rows from top to bottom present results in F070W to F444W filters. The median and upper and lower $1\sigma$ (84th and 16th percentiles $-$ median) of distributions of $\chi^2_{\nu}$ are shown at the upper-right corner.}
\label{fig5}
\end{figure*}

\begin{figure*}[t]
\centering
\includegraphics[width=\textwidth]{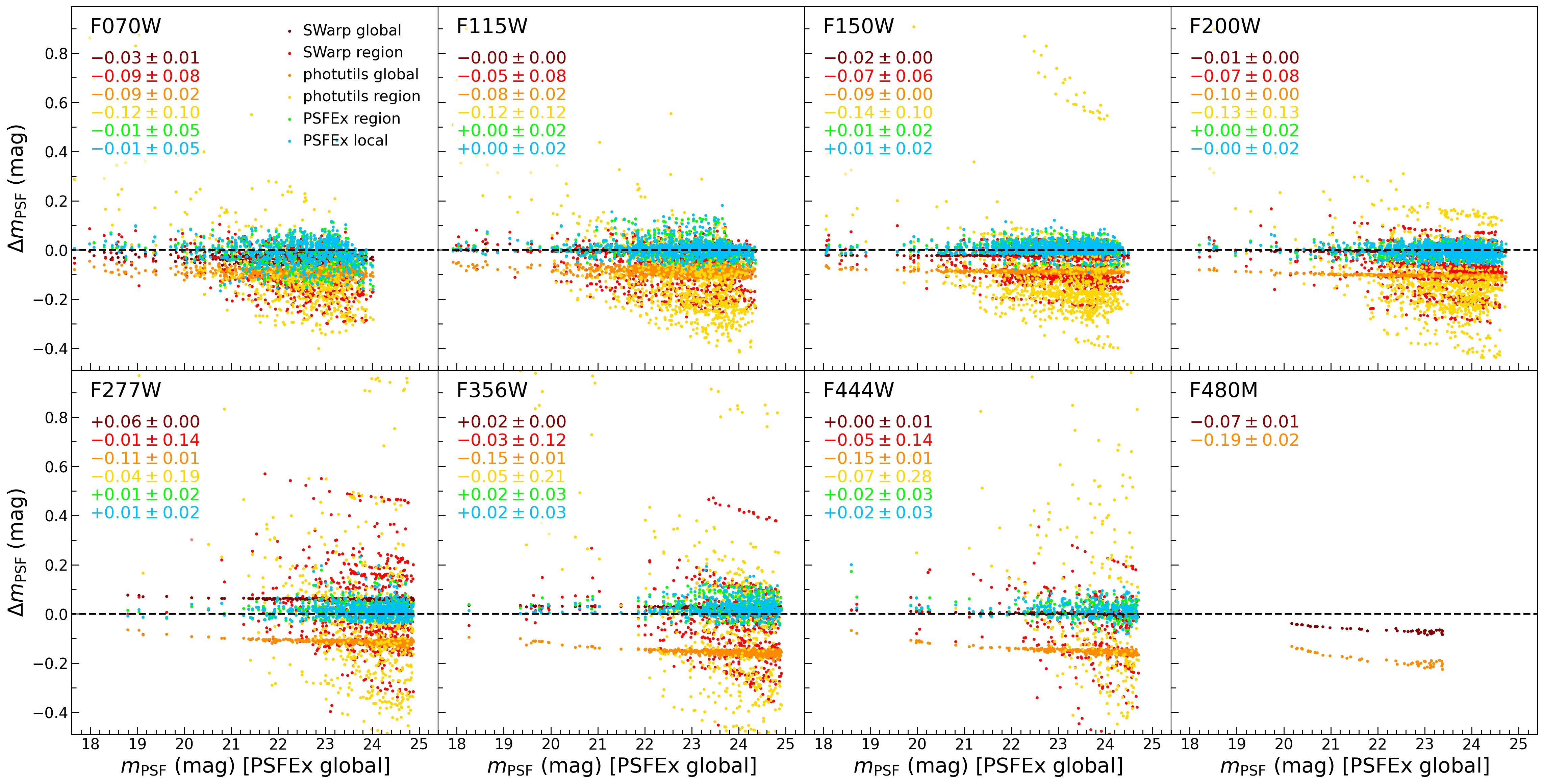}
\caption{Magnitude differences ($\Delta=$ PSF model $-$ \texttt{PSFEx} {\em global}) of point sources measured using different PSF construction methods versus magnitudes ($m_{\rm PSF}$) measured using \texttt{PSFEx} {\em global}. Colors are the same as Figure~\ref{fig5}. Median differences and standard deviations of $\Delta m_{\rm PSF}$ are shown at the top-left corner. Uncertainties of $m_{\rm PSF}$ are $\lesssim0.01$ mag and are thus not shown. Due to $<5$ point sources available in any region of the F480M image, we only show comparison for {\em global} PSF models in that filter. Combined with Figure~\ref{fig7}, our results indicate that (1) the quality of the PSF models from \texttt{SWarp} and \texttt{photutils} can be significantly affected by background; (2) PSF variation does not significantly affect flux measurements for point sources. See Section~\ref{sec3.2.2} for details.}
\label{fig6}
\end{figure*}

\begin{figure*}[t]
\centering
\includegraphics[width=\textwidth]{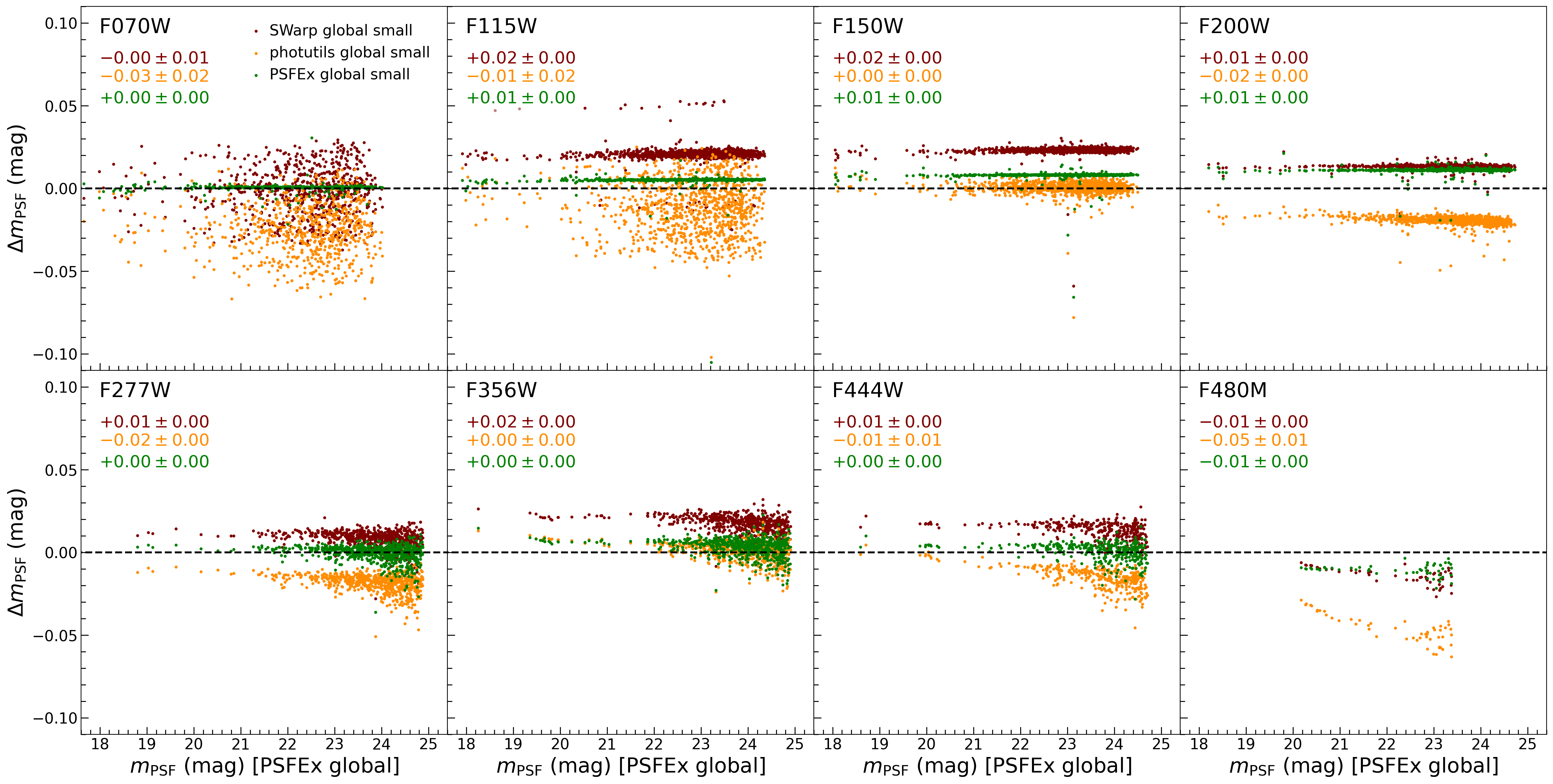}
\caption{Same as Figure~\ref{fig6} but with the minuend in the $\Delta m_{\rm PSF}$ substituted by $m_{\rm PSF}$ obtained from PSF models with half its original size, to reduce the contamination from imperfect background subtraction.}
\label{fig7}
\end{figure*}

\subsection{The Effect of PSF Spatial Variations on the Measurements of Point Sources}\label{sec3.2}

To investigate the effect of PSF spatial variations on the measurements of point sources, we fit their images with \texttt{gloabl} and the corresponding {\em region} PSF models constructed from the three methods using \texttt{galfit} \citep{Peng+2002AJ, Peng+2010AJ}. In particular, as \texttt{PSFEx} can provide a PSF model at any location based on interpolation, we also include a {\em local} PSF model for each source using its position on the mosaic. 

\subsubsection{Light Distribution of Point Sources}\label{sec3.2.1}

Figure~\ref{fig5} shows the normalized probability density distribution of reduced $\chi^2$, which is an indicator of goodness of the fit in terms of the light profile. We divide the sources into 4 categories: (a) 10 brightest sources, (b) sources with magnitude between 0th and 5th percentiles, (c) sources between 5th and 16th percentiles, and (d) sources between 16th and 100th percentiles. We also limit our tests to regions that contain at least 5 point sources used for PSF construction to avoid the problem of ``fitting one object with itself''. All PSF models provide overall good fits ($\chi^2_{\nu}\approx 1-2$) to the majority (84\%) of the sources (Figure~\ref{fig5}d). However, the performance of different PSF models tends to vary as source becomes brighter. For three {\em global} PSF models, \texttt{PSFEx} performs slightly better than \texttt{SWarp}, and both are noticeably better than \texttt{photutils}. For {\em region} PSF models, \texttt{photutils} provides the least satisfying results among 7 types of PSF models, likely due to the relatively low quality PSF models built based on a small number of objects (see Figure~\ref{fig2} for examples of PSF models using 11--26 objects in the F070W filter). {\em region} PSF models from \texttt{SWarp} perform similarly well as {\em global} PSF models in F070W, F115W, and F150W, but slightly worse in F200W and the three LW filters. {\em region} and {\em local} PSF models from \texttt{PSFEx} have slightly improved performances compared to {\em global} \texttt{PSFEx} in F070W and F115W, likely due to the relatively larger PSF spatial variations in these bands that are properly accounted for with the {\em region} and {\em local} models. Based on these results, we conclude that the three PSF models ({\em global}, {\em region}, and {\em local}) from \texttt{PSFEx} provide the best performance among 7 PSF models in describing the light distribution of point sources. 

However, we note that although \texttt{PSFEx} constructs PSF models by modeling multiple point sources simultaneously, its high order vector images may still have relatively poor SNR due to limited number of objects located close to the edges (Figure~\ref{fig2}). Therefore, we check the performance of {\em region} PSF models from \texttt{PSFEx} using the median $\chi^2_{\nu}$ for all point sources in each region. We find that {\em region} PSF models for regions that are close to the edges perform equally well as those for inner regions, albeit with lower SNR. This demonstrates the overall robustness and reliability of \texttt{PSFEx} in modeling the spatial variations of the PSF.

\subsubsection{PSF Magnitude of Point Sources}\label{sec3.2.2}

Figure~\ref{fig6} compares the magnitude of point sources measured using different PSF models. We use PSF magnitude ($m_{\rm PSF}$) measured using the {\em global} \texttt{PSFEx} PSF model as the reference, given its high SNR in the PSF wings and overall good performance in describing the light distribution of point sources (Section~\ref{sec3.2.1}). We find that $m_{\rm PSF}$ measurements from three \texttt{PSFEx} PSF models are very consistent with each other, with $|\Delta m_{\rm PSF}| \lesssim0.02$ mag in systematic offset and $\leq0.05$ mag in random scatter across all the filters. Both PSF spatial variation and larger noise toward the fainter end of sources could contribute to the observed total scatter. The largest scatter observed in the F070W filter is most likely driven by its significant PSF spatial variations (Table~\ref{table1}). The systematic differences, although rather mild, could result from the uneven distribution of sources in the FoV. In summary, our results suggest that the spatial variations of the PSF have a mild ($\lesssim0.05$ mag or 5\%) effect on measuring the magnitude of point sources. 

For {\em global} PSF models from the other two methods, we find that they are in general good agreements, albeit with consistently (scatter $\lesssim0.02$ mag) larger systematic differences with respective to the reference magnitudes. Noticeably, the offsets are higher for \texttt{photutils} PSF models ($|\Delta m_{\rm PSF}| \lesssim0.19$ mag) than \texttt{SWarp} PSF models ($|\Delta m_{\rm PSF}| \lesssim0.07$ mag). The consistence is much worse for {\em region} PSF models from \texttt{SWarp} ($|\Delta m_{\rm PSF}|$ up to 0.09 mag and scatter up to 0.14 mag) and \texttt{photutils} ($|\Delta m_{\rm PSF}|$ up to 0.15 mag and scatter up to 0.28 mag). We find that the systematic magnitude offset is mainly driven by inadequate background subtraction. As we are using a large cutout image size (135$\times$135 pixels for SW filters and 203$\times$203 pixels for LW filters) for PSF construction, the outskirt of the majority of the sources is dominated by background and thus could be affected by imperfect background subtraction. To test this hypothesis, we reduce the size of the image of PSF models by half and re-normalize it by subtracting the median value of pixels close to the edges (5 pixel wide). As shown in Figure~\ref{fig7}, the substantial magnitude offsets ($|\Delta m_{\rm PSF}|\lesssim0.19$) for {\em global} PSF models are significantly reduced to $|\Delta m_{\rm PSF}|\lesssim0.03$ mag for both \texttt{SWarp} and \texttt{photutils} in all filters except F480M. The residual systematic offsets could arise from subtle differences in FWHM (as a proxy of the PSF model profile) and minor over- (under-)estimation of background noise. Similarly, {\em region} PSF models, in particular those of \texttt{photutils}, are more likely to suffer from the factors mentioned above and exhibit a larger degree of inconsistency, mainly due to fewer sources available for PSF construction in individual regions. The small offsets ($\leq0.01$ mag) between \texttt{PSFEx} PSF models of two sizes in all the filters suggest that this method is insensitive to background variations and subtraction.

\subsection{Conclusions on PSF Characterization and Recommendations}

Our results show that PSF models constructed using point sources in the drizzled mosaics with a pixel scale of 0\farcs03 pixel$^{-1}$ and \texttt{pixfrac=1} in eight NIRCam filters (F070W, F115W, F150W, F200W, F277W, F356W, F444W, and F480M) have FWHM $\gtrsim 2$ pixels. The NIRCam PSF has significant spatial variations across the FoV, with maximum and RMS variations of FWHM decreasing from $\sim20$\% and $5$\% in F070W to $\sim5$\% and 0.6\% in F444W. Different filters have diverse variation patterns with extreme FWHMs being predominately found close to the edges. Neglecting PSF spatial variations would have minor impact on the magnitude of point sources across the FoV ($\leq0.02$ mag systematic offset and $\leq0.05$ mag scatter), depending on the required photometric accuracy.

We recommend the use of \texttt{PSFEx} to construct the PSF model, as well as modeling its spatial variations. It can deliver PSF models with higher SNR toward larger radii, with better performance in recovering the light profile and magnitude of point sources compared to the other two methods. If a sufficient number of point sources are available, it is recommended to construct {\em region} or {\em local} PSF models. Otherwise, adopting the {\em global} PSF model still provides satisfactory results under most circumstances, with an average $\lesssim0.02$ mag systematic offset and $\lesssim0.05$ mag random scatter. For the use of the other two methods (\texttt{SWarp} and \texttt{photutils}) to construct PSF models, adopting a smaller PSF model size would significantly mitigate the effects of background fluctuations and imperfect background subtraction. 

\subsection{Future Work}

As our analysis is based on combined mosaics, the characteristics of the PSF and its spatial variations presented here are subject to the specific configuration of observations. Observations with various dither patterns and steps, number of exposures, and drizzling parameters (e.g., output pixel size, \texttt{pixfrac}, and flux distribution kernel) are required to fully explore their effects on the geometry and spatial variation of the PSF. 

Tests on mock NIRCam images show that smaller output pixel scale and \texttt{pixfrac} would produce PSFs with smaller FWHMs \citep{Bagley+2022CEERS_data_reduction}. More experiments on real observations are needed to determine the best parameters to use in producing the final images. On the other hand, PSF can vary with time and thus lead to field-to-field variations. \citet{Nardiello+2022MNRAS} find that the temporal variations of NIRCam PSF in the F090W filter are on the level of $3-4$\% using images from calibrated exposures (prior to drizzling) from two observations separated by about one month. This level of temporal variation is significant and comparable to the RMS spatial variations across the FoV in the F070W filter (4.7\%; Table~\ref{table1}). Extending our work to observations at different epochs would help us understand better the temporal variation of the PSF. Moreover, since most deep fields by design targeting at high galactic latitude, where few bright stars are available for PSF model construction, the effect of the number of usable point sources on the quality and properties of the PSF also merits further investigation. Last but not least, NIRCam employs Hawaii-2RG HgCdTe detectors, which are found to have a brighter-fatter effect \citep[e.g.,][]{Plazas+2018PASP}. A comprehensive and quantitative analysis of this effect on the FWHM and profile of observed NIRCam PSFs would be important for studies in crowded fields such as globular clusters.

\section{Implications on the AGN-Host Image Decomposition}\label{sec4}

For our work, the main motivation of PSF characterization of NIRCam imaging is to assess the reliability of host measurements in unobscured AGNs, or their high-luminosity counterparts, quasars, using imaging decomposition techniques. Below we apply our methodology of PSF model construction discussed above to simulated AGN+host images, and investigate the recoverability of host properties.


To achieve this goal, we generate mock AGN+galaxy images with a central point source (AGN) and a \sersic\ profile (host galaxy), convolved with the measured {\em global} PSF model from \texttt{PSFEx}. The centers of the AGN and the host are tied together in the decomposition with zero offset. These simulated systems, although oversimplified, provide critical information on the true uncertainties and biases of recovered host properties that are often underestimated from fitting \citep[e.g.][]{Haussler+2007ApJS, van_der_Wel2014ApJ, Zhuang&Ho2022ApJ}. For simplicity, we do not consider more complex substructures of the host galaxy, which could lead to additional systematics \citep{Zhuang&Ho2023}. 

We generate mock systems across a broad range of parameters. We consider AGN magnitude ($m_{\rm AGN}$) from 16.5 to 30.0 mag in steps of 0.5 mag; AGN-to-host ratio from 0.1 to 10 in steps of 0.2 dex, which corresponds to a step of 0.5 mag; effective radius ($R_e$) from 4 to 64 pixels (0\farcs12--1\farcs92) in steps of 0.3 dex, which covers the range of $\sim1-10$ kpc at $z=1-7.5$; \sersic\ index $n$ = 0.5, 1, 1.5, 2, 3, 4, and 6; axis ratio between major- and minor-axis $q$ = 0.3, 0.6, and 0.9; a fixed position angle of $45\degree$. For the convenience of applying our results to other fields with different exposure times (hence different depths), we measure the $5\sigma$ point source depth and $3\sigma$ surface brightness (SB) depth, where $\sigma$ is the standard deviation of 50 flux measurements randomly distributed in the background image within an aperture of radius = PSF FWHM {(for point sources)} and 0\farcs5 (for extended sources), respectively. The measurements for point sources are then corrected for flux loss outside the aperture to determine the final point-source depth. The results are presented in Table~\ref{table2}. 

In total, we generated 32,340 mock AGN+galaxy images in each of the 7 wide filters (except F480M) with a cutout size of $6\times R_e$ to enclose most of the emission from the host galaxy. Two types of noise are added to the mock images: (1) Poisson noise for the source and (2) background noise. For the background noise, we use real images from which the PSF models are constructed. Our final background cutout contains no bright sources with only $\sim0.3$\% of pixels having fluxes larger than $5\times$ background standard deviation. The original cutout has a size of $\sim180$ pixel (5\farcs4) and is padded in a $3\times3$ array to construct the final background image. 

We fit images of mock AGNs with four sets of fitting scenarios using \texttt{galfit}: (1) the fiducial model with the same PSF model as the one used to generate the mock images ({\em global} \texttt{PSFEx}); during the decomposition, the centers of the AGN and \sersic\ components are tied together without offset; (2) same as (1) but with a broader PSF model; (3) same as (1) but with a narrower PSF model; (4) same as (1) but allow the centers of AGN and \sersic\ to offset from each other. For all four fitting scenarios, we set limits of fitting parameters: magnitude to 10--35 mag, $R_e$ to 0.5--100 pixels, $n$ to 0.3--8, $q$ to 0.1--0.99, and center position constrained to within $\pm1$ pixel of its input value for the first three sets and $\pm3$ pixels for the last set. The fitting results for all mock images are made available to the public in the \href{https://github.com/mingyangzhuang/JWST-NIRCam-Data-Product}{github}.

Fitting scenarios (1) and (4) will evaluate the recoverability of host properties as functions of input parameters assuming a perfect PSF model, but with different fitting methodologies (whether or not tie together the centers of the two components). Fitting scenarios (2) and (3) will evaluate additional uncertainties from PSF mismatches. When using broader or narrower PSF models, we adopt two {\em region} PSF models with the largest and smallest FWHMs to illustrate the maximum variations in each filter -- they differ from the fiducial PSF FWHM by $\sim-7$\% and $+12$\% in the F070W and $\sim-2$\% and $+1$\% in the F444W filter. For illustration purposes, we show the results for host galaxies with $19\leq$ \msersic $\leq27.5$ mag in the F150W filter in the following part of Section 4. This filter displays a moderate level of spatial variations among all filters (Table~\ref{table1}), so that the results presented in Section~\ref{sec4.2} can be considered a typical case for a given filter. 

\begin{deluxetable}{cccc}
\caption{Properties of PSF Models \label{table2}}
\tabletypesize{\small}
\tablehead{
\colhead{Filter} & \colhead{$5\sigma$ PS depth} & \colhead{$3\sigma$ SB depth} & \colhead{Flux loss correction}\\
\nocolhead{} & \colhead{(mag)} & \colhead{(mag arcsec$^{-2}$)} & \nocolhead{}\\
\colhead{(1)} & \colhead{(2)} & \colhead{(3)} & \colhead{(4)}
}
\startdata
F070W & 26.42 & 24.94 & 0.329\\
F115W & 26.67 & 25.31 & 0.327\\
F150W & 26.83 & 25.28 & 0.332\\
F200W & 27.20 & 25.58 & 0.355\\
F277W & 26.82 & 26.19 & 0.363\\
F356W & 26.72 & 26.32 & 0.365\\
F444W & 26.48 & 26.29 & 0.382\\
\enddata
\tablecomments{Col. (1): Filter name. Col. (2-3): $5\sigma$ point source (PS) depth and $3\sigma$ surface brightness (SB) depth, where $\sigma$ is the standard deviation of 50 flux measurements randomly distributed in the background image within an aperture of radius = PSF FWHM and 0\farcs5, respectively. Measurement for PS is corrected using value in Col. (4) for fluxes outside the aperture.}
\end{deluxetable}

\begin{figure}[t]
\centering
\includegraphics[width=0.5\textwidth]{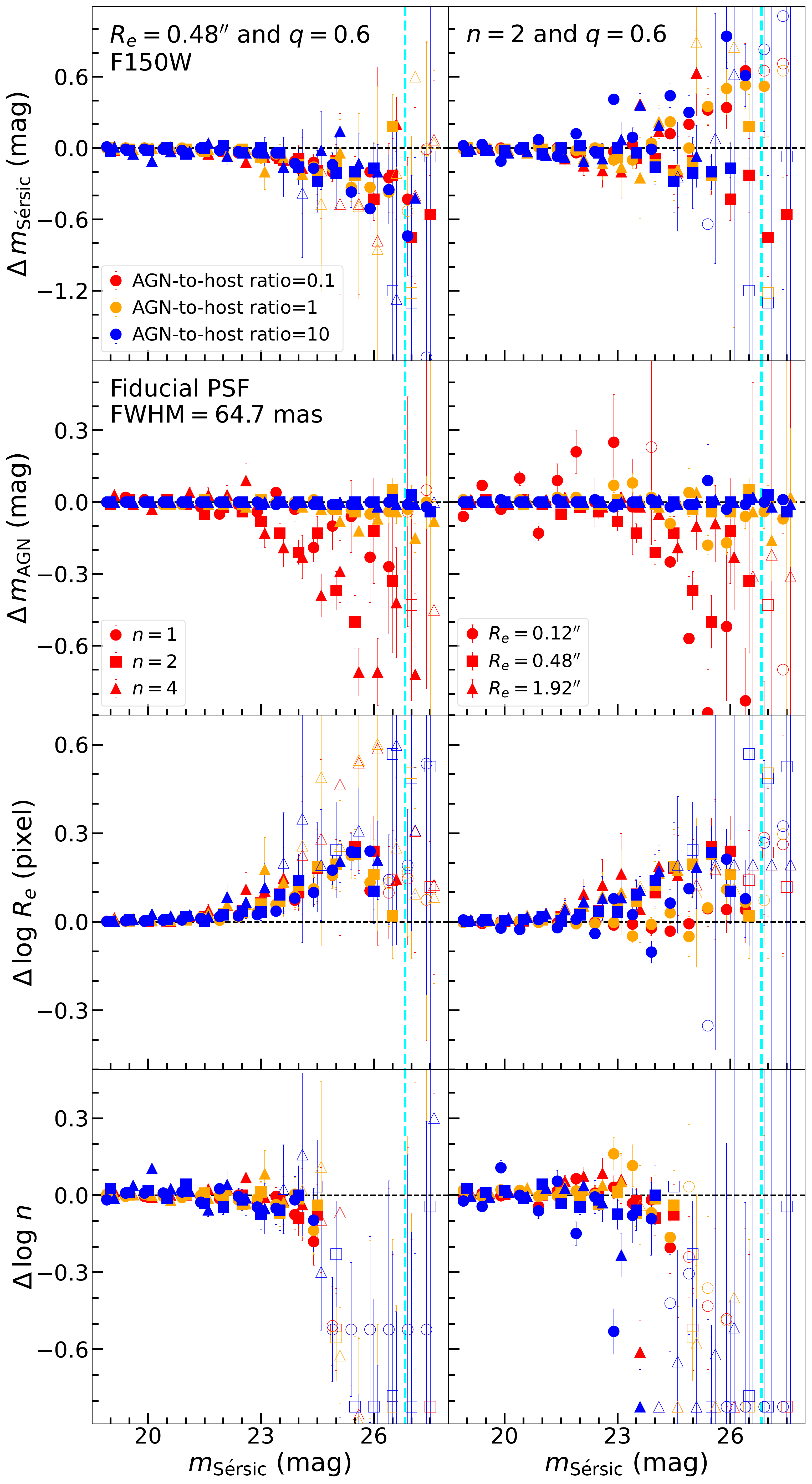}
\caption{Differences ($\Delta$ = output $-$ input) between the best-fit results and the input values for \msersic\ (first row), $m_{\rm AGN}$ (second row), $R_e$ (third row), and \sersic\ index $n$ (fourth row) versus \msersic\ using the perfect PSF model. The left column shows mock AGNs with fixed $R_e$ and $q$ but different $n$, while the right column shows mock AGNs with fixed $n$ and $q$ but different $R_e$. Red, orange, and blue colors represent objects with AGN-to-host ratio $=$ 0.1, 1, and 10, respectively. Open symbols represent objects with measurement $<3\times$ its reported uncertainty. The vertical cyan dashed line indicates the $5\sigma$ point source depth. {Error bars represent uncertainties for individual objects as reported by \texttt{galfit}.}}
\label{fig8}
\end{figure}

\subsection{Reliability of Host Measurements}\label{sec4.1}

Figure~\ref{fig8} compares the recovered parameters of mock AGNs with the input values, as a function of \msersic\ using the {\em global} PSF model from \texttt{PSFEx} in the F150W filter. This test represents the situation where the PSF model is perfect (i.e., the mock images are generated using the same PSF model). We find overall good consistency between the recovered parameters and the input values for bright host galaxies. As host galaxies become fainter, the recovered host properties tend to show larger measurement uncertainties, larger offset (scatter) compared with input values, and a larger fraction of objects have failed measurements (i.e., measurement/uncertainty$<3$; open symbols). In addition, the consistency between recovered host parameters (\msersic, $R_e$, and $n$) and the input values degrades toward higher AGN-to-host ratio, while the recovery of $m_{\rm AGN}$ becomes worse toward lower AGN fractions. No obvious differences are found among systems with different $q$ values. As long as \msersic\ is successfully measured (measurement/uncertainty$\geq 3$), $q$ can be estimated correctly as well, albeit with larger scatter toward higher AGN-to-host ratio (e.g., fractional scatter increases by a factor of $\sim3$ at $q=0.6$, from 3\% for AGN-to-host ratio=0.1 to 10\% for AGN-to-host ratio=10). 

At fixed $R_e$, we find systematic overestimation of the host flux (smaller \msersic) due to a combination of overestimated $R_e$ and underestimated $n$. $m_{\rm AGN}$ tends to be underestimated (flux overestimated) when host is more concentrated (larger $n$). At fixed $n$, both small and large $R_e$ can lead to slight underestimation of host flux and larger scatter in all four parameters presented here. 

Among the three host parameters, \msersic\ has the highest fraction of successful measurements, followed by $R_e$ and $n$. The exact success rate depends on the AGN-to-host ratio and host structure. Galaxies with higher $n$ and both small and large $R_e$ are more likely to have unsuccessful measurements. \msersic\ and $R_e$ in objects with low AGN-to-host ratios, small $n$ and moderate $R_e$ can be successfully measured down to the $5\sigma$ point source depth. However, no successful measurement can be obtained at \msersic$>25$ mag ($\sim2$ mag above the $5\sigma$ point source depth) for $n$, with little dependence on the AGN-to-host ratio. 

The deviations from input values ($\Delta$) are larger than the formal measurement uncertainties (error bars in Figure~\ref{fig8}) for the majority of cases where successful measurements are obtained. For example, median $|\Delta|$/uncertainty of \msersic\ increases from $\sim1.2$ at AGN-to-host ratio=0.1 to 5.0 at AGN-to-host ratio=10, with extreme values as high as $\sim100$. This urges the need of generating object-tailored mock AGNs with input parameters from the best fit to evaluate systematic biases and to obtain more realistic measurement uncertainties.

\begin{figure}[t]
\centering
\includegraphics[width=0.5\textwidth]{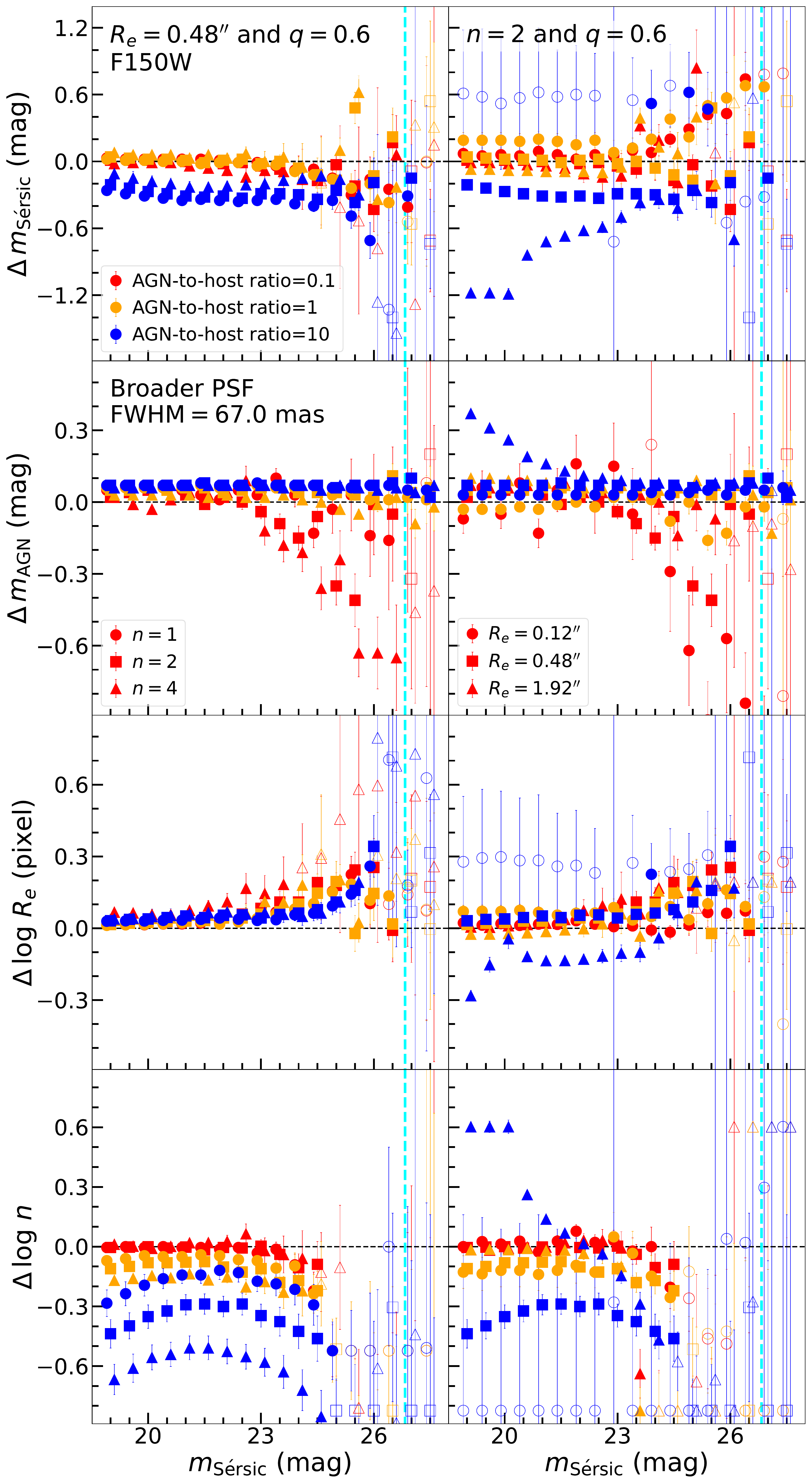}
\caption{Same as Figure~\ref{fig8} but for the results using a broader PSF model in the decomposition.}
\label{fig9}
\end{figure}

\begin{figure}[t]
\centering
\includegraphics[width=0.5\textwidth]{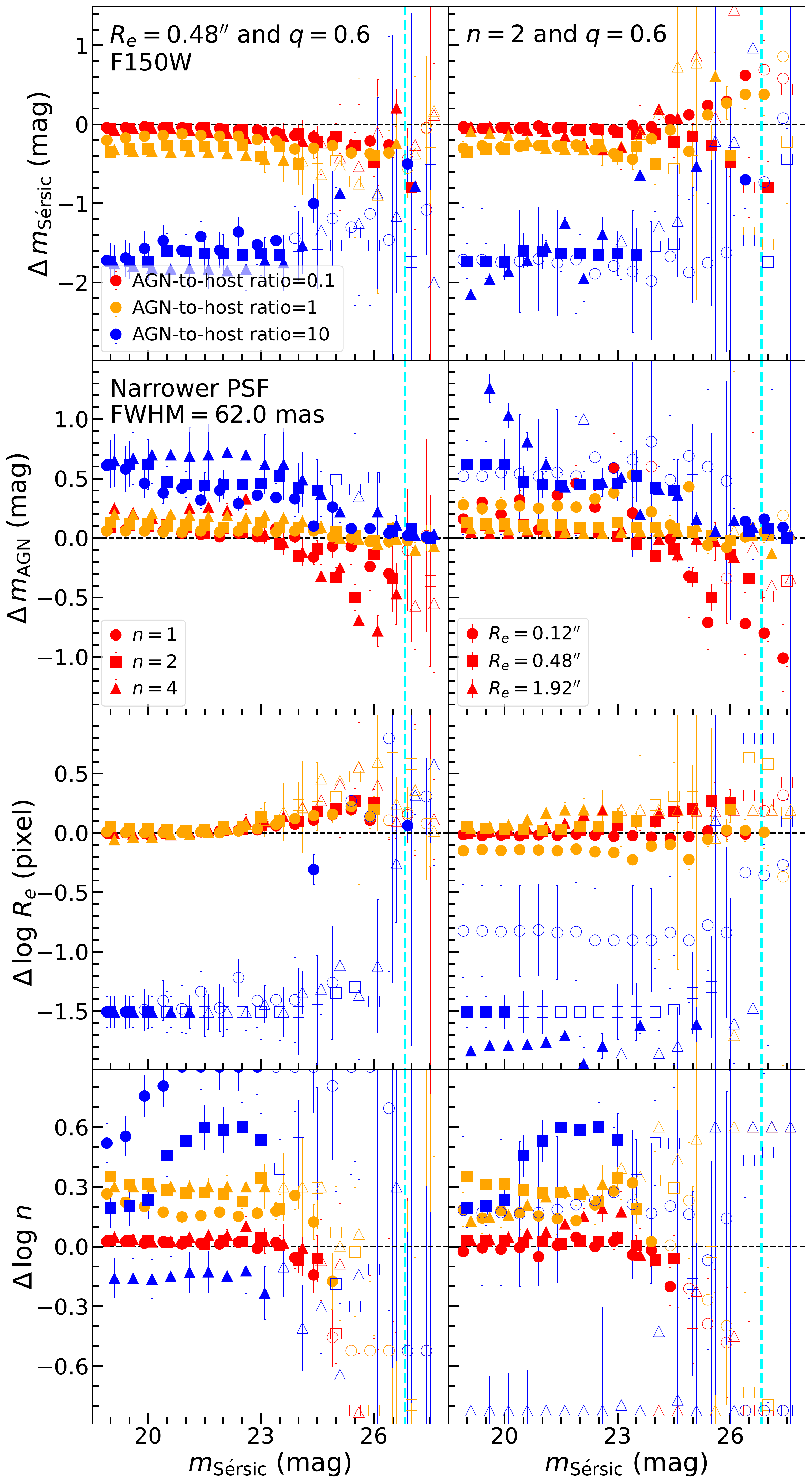}
\caption{Same as Figure~\ref{fig8} but for the results from the narrower PSF model.}
\label{fig10}
\end{figure}

\begin{figure}[t]
\centering
\includegraphics[width=0.49\textwidth]{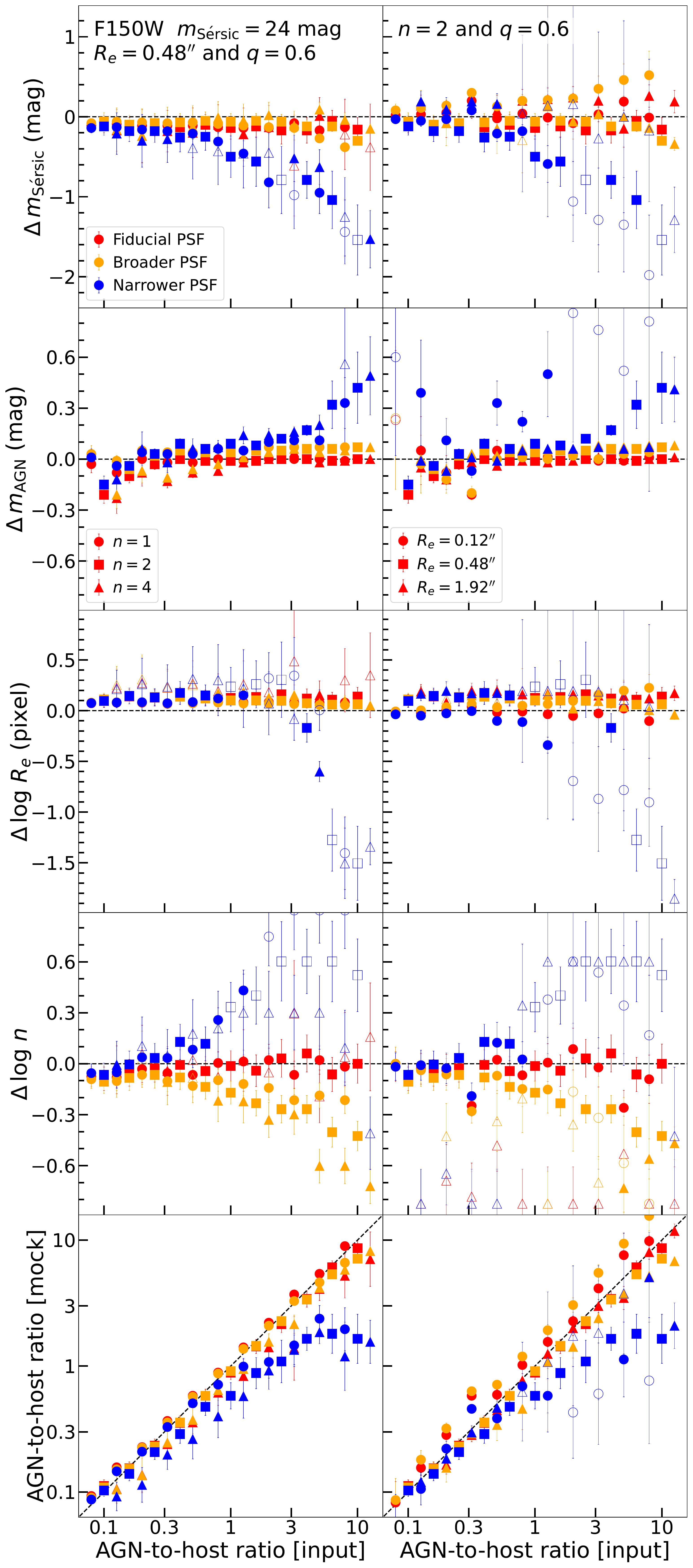}
\caption{Same as Figure~\ref{fig8} but for differences of properties versus AGN-to-host ratio in mock AGNs with \msersic$=24$ mag. Symbols are the same as Figure~\ref{fig8} with red, orange, and blue colors representing results from the fiducial, broader, and narrower PSF models, respectively. }
\label{fig11}
\end{figure}

\begin{figure*}[t]
\centering
\includegraphics[height=0.4\textwidth]{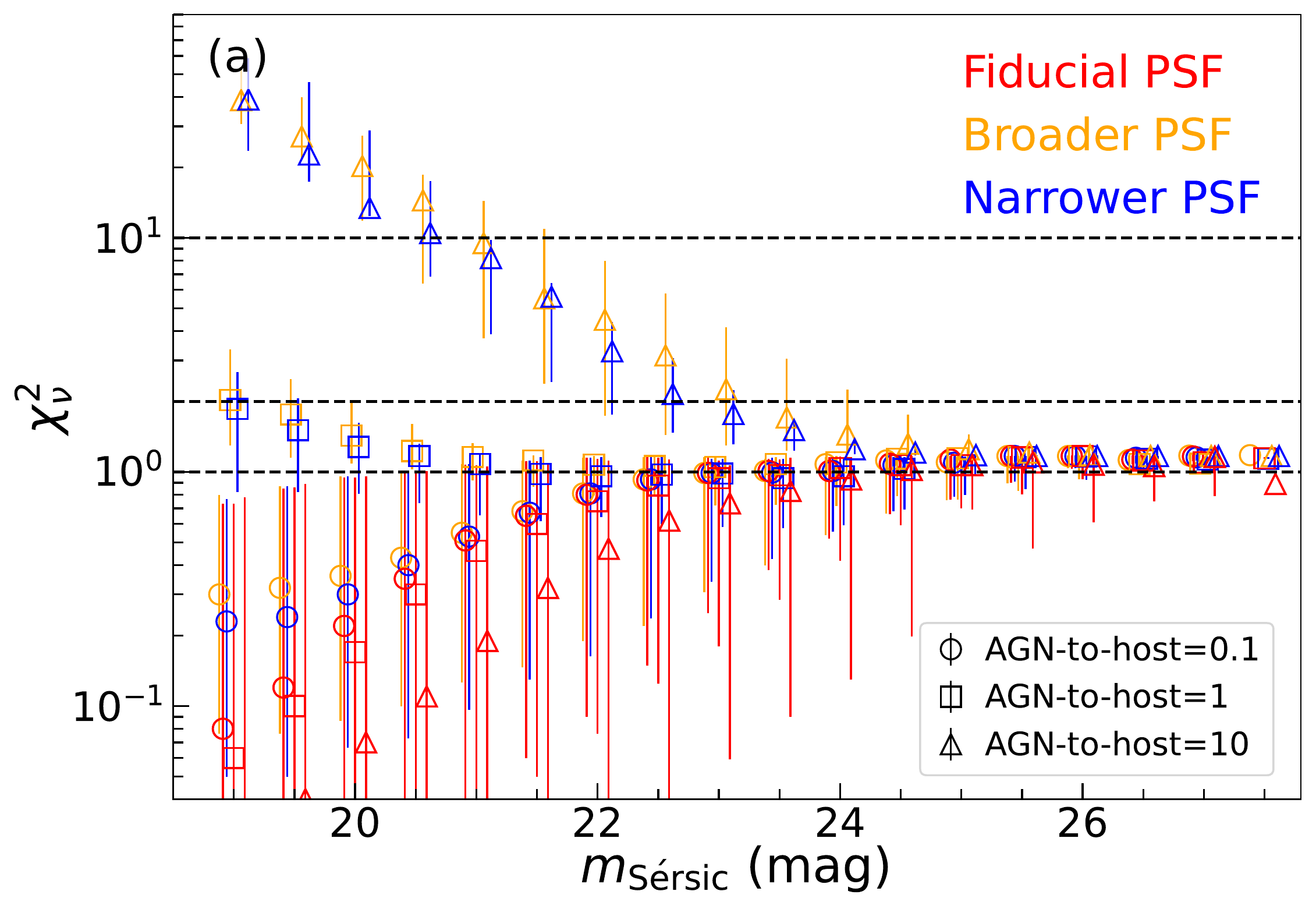}
\includegraphics[height=0.4\textwidth]{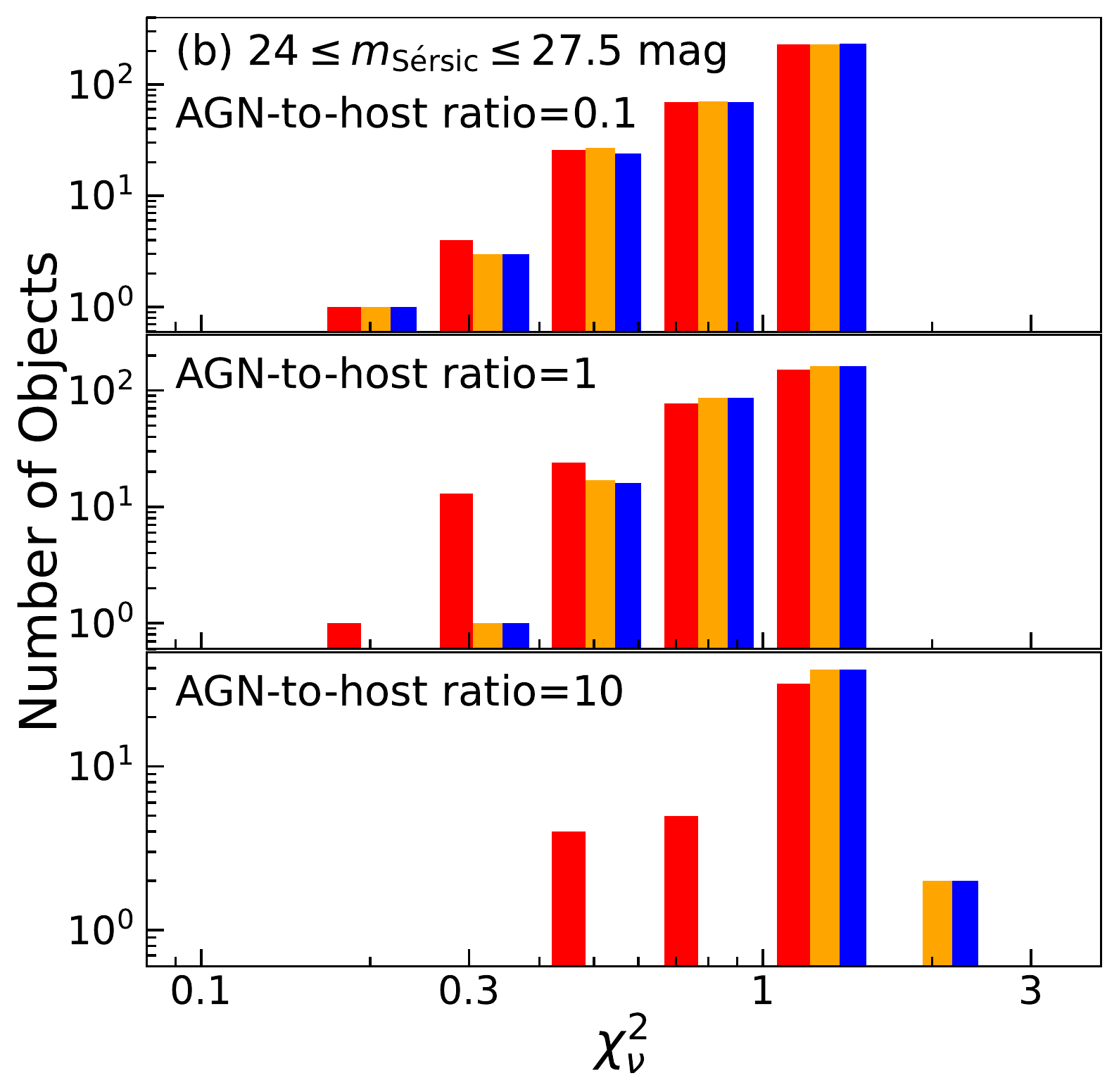}
\caption{{(a): Median $\chi_{\nu}^2$ of decomposition to mock AGNs with successful \msersic\ measurement from \texttt{galfit} versus input \msersic\ for the fiducial (red), broader (orange), and narrower (blue) PSF models. Circles, triangles, and squares represent objects with AGN-to-host ratio = 0.1, 1, and 10, respectively. 16th and 84th percentiles are indicated with errorbar. Input \msersic\ are shifted by small values to mitigate symbol overlapping.} {(b): Histograms of $\chi_{\nu}^2$ for objects with $24\leq$ (\msersic/mag) $\leq27.5$ AGN-to-host ratio = 0.1 (top), 1 (middle), and 10 (bottom).}}
\label{fig12}
\end{figure*}

\subsection{The Effect of PSF Mismatch on Host Measurements}\label{sec4.2}

We now investigate the impact of PSF mismatch on the recovery of host parameters. Figure~\ref{fig9} and \ref{fig10} show the results from a broader (3.6\% broader in the FWHM selected from region 6) and a narrower (4.2\% narrower in the FWHM selected from region 31) PSF model in the F150W filter, respectively. 

Recovered parameters from the broader and narrower PSF models generally follow the same trends as those using the perfect PSF model for mock AGNs with AGN-to-host ratio=0.1 and 1, albeit with slightly larger scatter and offsets. Significantly larger offsets are observed in AGN-dominated systems (AGN-to-host ratio=10). 

We find that although the total fluxes of host galaxies are more likely to be overestimated (smaller \msersic) for both mismatched PSF models, the reasons are different. The broader PSF model tends to underestimate $n$, predicting more extended host galaxies. This behavior is expected since the emission from the central part of the host is mistaken as part of the unresolved emission from the AGN with the broader PSF model, causing the fitted host to be more extended. Since \sersic\ models with larger $n$ contain a larger fraction of emission outside $R_e$, the more extended emission at large radii coupled with the true $n$ from the host galaxy will upscale the overall flux from the \sersic\ model, when the fitted $n$ is underestimated using a broader PSF model.

On the other hand, the narrower PSF model tends to underestimate $R_e$ and overestimate $n$, predicting more concentrated and smaller host galaxies. Therefore, with the narrower PSF model, the emission from the AGN is split into contributions from a narrower point component and an extended component. Consequently, AGN flux is underestimated and host flux is overestimated for the narrower PSF model. For systems with AGN-to-host=10 (i.e., AGN-dominated systems), the underestimation of AGN flux can reach as much as $\sim0.6$ mag, and the overestimation of host flux can reach as much as 1.8~mag! 

Moreover, in systems with AGN-to-host=10 and successful \msersic\ measurements from the narrower PSF model, the measured $q$ values are generally around $\sim0.6$ regardless of their input values, making it difficult to estimate the inclination angle of the host.

The structure of the host galaxy plays a more important role on the host parameter recovery when the PSF model is mismatched. Measurements are more likely to fail (measurement/uncertainty$<3$) in small host galaxies. The broader PSF model tends to predict smaller \msersic, smaller $R_e$, and higher $n$ toward larger host galaxies (larger $R_e$) and lower $n$ toward more concentrated host galaxies (larger $n$). The narrower PSF model follow the same trends as the broader PSF model except for the relation of $n$ and galaxy size (no meaningful trend due to no successful measurements for hosts with $R_e=$0\farcs12 and 1\farcs92). 

For the measurement success rate, the broader PSF model generally produce similar results as the fiducial PSF model. The only obvious difference is for the host galaxies with small $R_e$. The measurement success rate is much lower for the narrower PSF model. For example, for host galaxies with AGN-to-host ratio=10, almost no successful measurements can be obtained for $R_e$ at \msersic$\gtrsim21.5$ mag ($\sim5.5$ mag above the point source depth). 

To better facilitate a detailed look at the effect of the AGN-to-host ratio, Figure~\ref{fig11} shows the differences of the parameters as a function of AGN-to-host ratio in mock AGNs with \msersic$=24$ mag. Results from the broader PSF model are similar to those from the fiducial PSF model, except for $n$ where gradual offset and scatter start to appear at AGN-to-host ratio$\gtrsim1$. However, the narrower PSF model would lead to noticeable offsets and scatters in all five parameters. {The AGN-to-host ratio tends to be underestimated in AGN-dominated systems (AGN-to-host ratio$\gtrsim1$), leading to a higher host fraction.} Almost no successful measurements of $R_e$ and $n$ can be obtained at AGN-to-host ratio$\gtrsim1$. \msersic\ can be significantly underestimated (hence host flux overestimated) at AGN-to-host ratio$\gtrsim1$. 

The goodness of fit can be assessed with the reduced $\chi^2$ ($\chi_{\nu}^2$). Figure~\ref{fig12}a presents the median $\chi_{\nu}^2$ from \texttt{galfit} decomposition of our mock images with successful \msersic\ measurements. For systems with AGN-to-host ratio=0.1 or 1, $\chi_{\nu}^2$ from \texttt{galfit} is generally $\lesssim1$ across the full \msersic\ range (\msersic$>19$ mag). Although results with the fiducial PSF model often have the lowest absolute values of $\chi_{\nu}^2$, we are unable to tell which PSF model performs the best, given the overall good performance of all three PSF models. On the other hand, when the host and the AGN are both well above the flux limit (e.g., \msersic$<21$ mag or $\sim6$ mag above the PS depth, and AGN-to-host ratio=10), the $\chi_{\nu}^2$ metric is a good parameter to select the best PSF model, since the S/N of both the AGN and host components is extremely high.

However, as the whole system approaches the imaging depth, $\chi_{\nu}^2$ values from the broader and narrower PSF models start to become comparable to that from the fiducial PSF model. At \msersic$\gtrsim23$ mag the $\chi_{\nu}^2$ values from three PSF models are all statistically acceptable for the fit. In fact, in some cases, the mismatched PSF models produce smaller $\chi_{\nu}^2$ values than the fiducial PSF model. We find that the distributions of $\chi_{\nu}^2$ of three PSF models in systems with $24\leq$ (\msersic/mag) $\leq27.5$ are quite similar (Figure~\ref{fig12}b), albeit with differences in the number of objects with successful \msersic\ measurements for different AGN-to-host ratios. Real systems have light profiles that are more complicated than a PSF + a single \sersic\ due to the presence of substructures (such as bulge, bar, lens, dust lane, and star-forming clumps), and the discriminating power of $\chi_{\nu}^2$ may be further weakened.

These tests on $\chi_{\nu}^2$ suggest that unless both the AGN and the host are much brighter than the imaging depth, a smaller $\chi_{\nu}^2$ does not necessarily mean the PSF model is correct. In fact, for most imaging applications where the host (or the full system) is often near the detection limit, PSF mismatch is a realistic concern and cannot be easily identified with the $\chi_{\nu}^2$ statistic. The systematic biases in the derived parameters due to PSF mismatches, as we demonstrated here, must be taken into account in the total error budget.


To summarize, PSF mismatch has significant effects on the derived host (and AGN) parameters from AGN-host image decomposition. Adopting a broader PSF model would lead to overestimated host fluxes with more extended host structures. Adopting a narrower PSF model could also lead to overestimated host fluxes but with more concentrated and smaller host structures. For a given object, adopting a narrower PSF model would introduce much worse systematic biases (e.g., underestimate \msersic\ by $\sim1.6$ mag) and lower measurement success rate compared to adopting a broader PSF model. The systematic biases caused by adopting a narrower PSF model are noticeable even in host dominated systems.

\begin{figure}[t]
\centering
\includegraphics[width=0.5\textwidth]{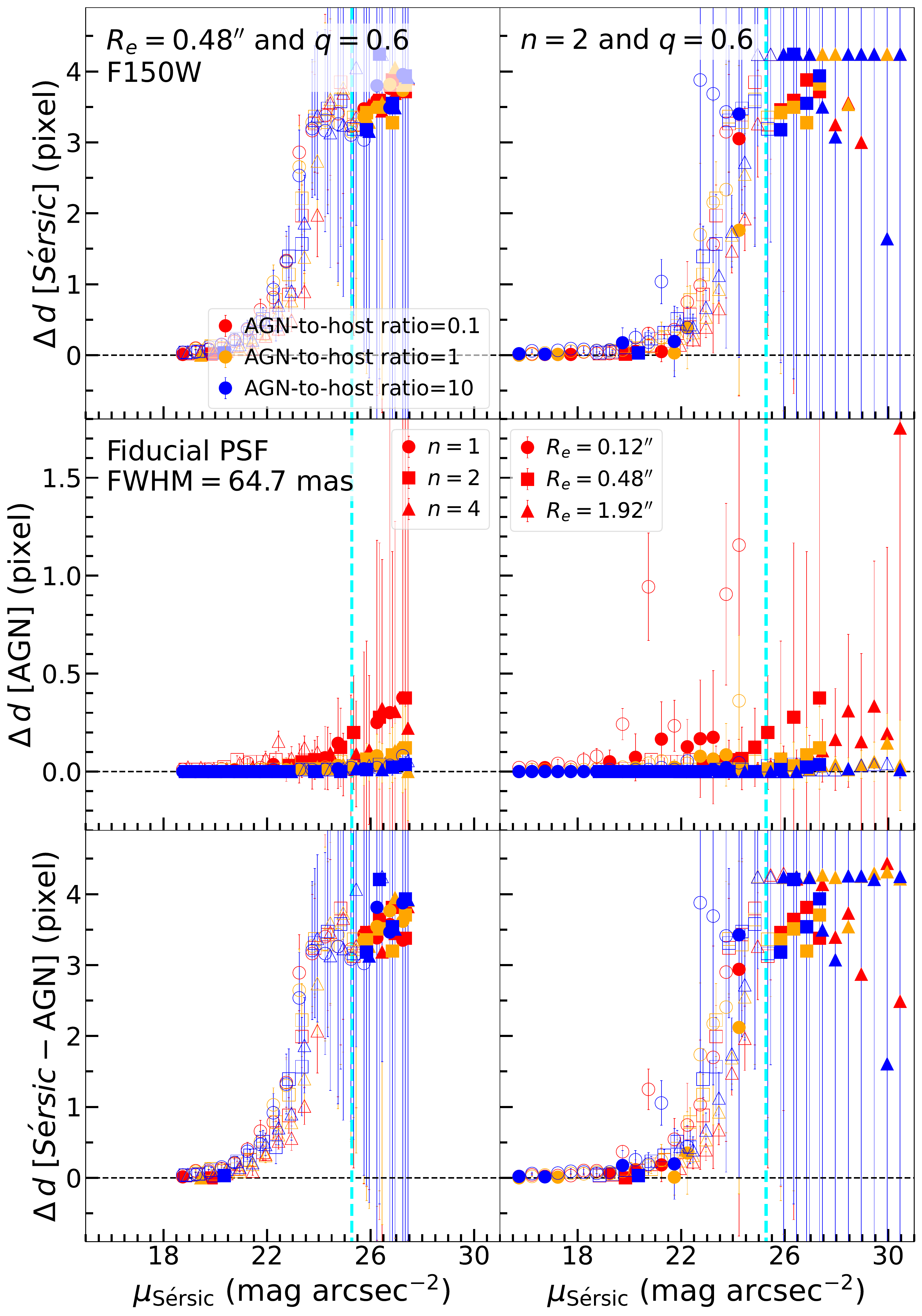}
\caption{Offsets of host center ($\Delta d$ [\sersic]; first row) and AGN center ($\Delta d$ [AGN]; middle row) between measurements in mock AGNs and input values, and offsets between center of AGN and host measured in mock AGNs ($\Delta d$ [\sersic$-$AGN]; bottom row) versus versus mean surface brightness of the \sersic\ model ($\mu_{S\acute{e}rsic}$). Mock AGNs are fitted with the fiducial PSF model but with centers of PSF and \sersic\ model untied. Color and symbols are the same as Figure~\ref{fig8}, but with open symbols indicating objects with offset larger than its uncertainty.}
\label{fig13}
\end{figure}

\subsection{Reliability of AGN-Host Offset Measurement}\label{sec4.3}

Figure~\ref{fig13} compares the differences of center measurements versus the mean surface brightness of the \sersic\ component ($\mu_{S\acute{e}rsic}$) between the fiducial fitting scenario and scenario 4, where in the latter case the centers of the AGN and host components are untied. $\mu_{S\acute{e}rsic}$ is defined as \msersic\ $+2.5\log(2 q\pi R_e^2)$. The factor of 2 is to account for the fact that half of the total emission is within the effective radius.

We find that one can easily measure an artificial offset between the AGN and its host galaxy centroid, mainly due to incorrect center measurements of host galaxies. The artificial offset generally follows the direction of the position angle of the host (i.e., 45\degree), randomly distributed on either side. The amount of offset increases rapidly with decreasing surface brightness and hits the boundary of $3\times \sqrt{2}$ pixel when the surface brightness of the host drops below the SB detection limit. The trends are similar in host galaxies with different structures. When considering all the sources with $19\leq$\msersic$\leq27.5$ mag, we find that offset measurements in 59--64\%, 28--40\%, and 11--21\% of objects are larger than 1$\times$, 2$\times$, and 3$\times$ the uncertainty reported by \texttt{galfit}, respectively. These percentage ranges encompass the AGN-to-host ratio range in our mock data (0.1--10). Excluding objects with $\mu_{S\acute{e}rsic}$ below the SB detection limit (associated with very large measurement uncertainty), offsets in 76--82\% (15--29\%) of the objects are larger than the reported $1\sigma$ ($3\sigma$) uncertainty. This fraction decreases toward larger AGN-to-host ratio due to larger position measurement uncertainties {for both the AGN and its host.} 

We note that the results above are only for idealized mock AGNs. With the presence of galactic substructures (including both axial symmetric and non-axial symmetric such as bulge, nuclear bar, spiral arms, lens, and dust lane) at the center of the galaxy, we expect more complicated trends in the measured artificial offset. The offsets can be as large as $\sim80$\%, 26\%, and 7\% of $R_e$ in objects with $R_e=$0\farcs12 (4 pixels), 0\farcs48 (16 pixels), and 1\farcs92 (64 pixels) when approaching the SB detection limit, respectively.

For completeness, no obvious differences in the global trends and scatter of the four parameters (\msersic, $m_{\rm AGN}$, $R_e$, and $n$) are found whether the centers of AGN and host are tied in the fit or not. 

\section{Applications of Our Methodology}\label{sec5}

As a demonstration of applying our method of optimal PSF modeling to study the properties of host galaxies of broad-line AGNs, we select a broad-line AGN SDSS1420+5300A (R.A. $=215.02333\degree$, Decl $=53.01020\degree$) at $z=1.646$, covered by Module A of NIRCam imaging in pointing 1 of the CEERS fields. This AGN has a BH mass ($M_{\rm BH}$) of $M_{\rm BH}= 10^{9.00 \pm 0.16} M_{\odot}$ from reverberation mapping \citep{Grier+2019ApJ}.

\subsection{PSF Model Construction}\label{sec5.1}

CEERS NIRCam observations have seven filters: F115W, F150W, F200W, F277W, F356W, F410M, and F444W. Six wide filters are considered here for the AGN-host decomposition. The early NIRCam observations are taken on June 21-22, 2022 in a coordinated parallel mode with MIRI imaging. We adopt the version 0.5 NIRCam imaging data product from CEERS team \citep{Bagley+2022CEERS_data_reduction}, which covers all four pointings (1, 2, 3, and 6) of June 2022 observations. Two types of observing configurations are adopted due to the use of different MIRI filters. For pointings 1 and 2, dither patterns of NIRCam are tied to the corresponding MIRI filters that are observed simultaneously, while for pointings 3 and 6, the same \texttt{3-POINT-MIRIF770W-WITHNIRCam} dither pattern is adopted. Therefore, we construct PSF models for this object using only point sources in the Module A of pointings 1 and 2. To model the spatial variations of the PSF, only a first order polynomial (linear) dependence provides usable PSF models due to the very limited number ($\sim20$) of point sources in the field. We construct three PSF models to investigate the differences of host properties: (1) a {\em local} PSF model constructed at the location of the AGN in the image; (2) a {\em global} PSF model; (3) a PSF model from (2) but broadened by a circular Gaussian kernel, so that the FWHM of the output PSF model is $1\sigma$ broader. Here $\sigma$ is the RMS fractional variation presented in Col. (7) of Table~\ref{table1}. 

There are several reasons for testing three PSF models: (1) the linear dependence of PSF model constructed locally at the AGN may not be accurate enough; (2) the {\em global} PSF model is representative of the median PSF model across the FoV of NIRCam; (3) in case the {\em global} PSF model underestimates the true PSF of the target, we slightly broaden it to mitigate the biases in the recovery of host properties, since a narrower fitting PSF can have severe consequences in the host recovery (Section~\ref{sec4.2}). We find that the {\em local} PSF models are generally narrower compared to the {\em global} ones except in the F150W filter. The FWHMs of our PSF models are slightly smaller than those reported in \cite{Finkelstein+2023ApJ}, which are constructed by stacking stars across all four pointings. We make our {\em global} PSF models for CEERS fields publicly available to the community in \href{https://github.com/mingyangzhuang/JWST-NIRCam-Data-Product}{github}.

\begin{figure*}[t]
\centering
\includegraphics[width=0.95\textwidth]{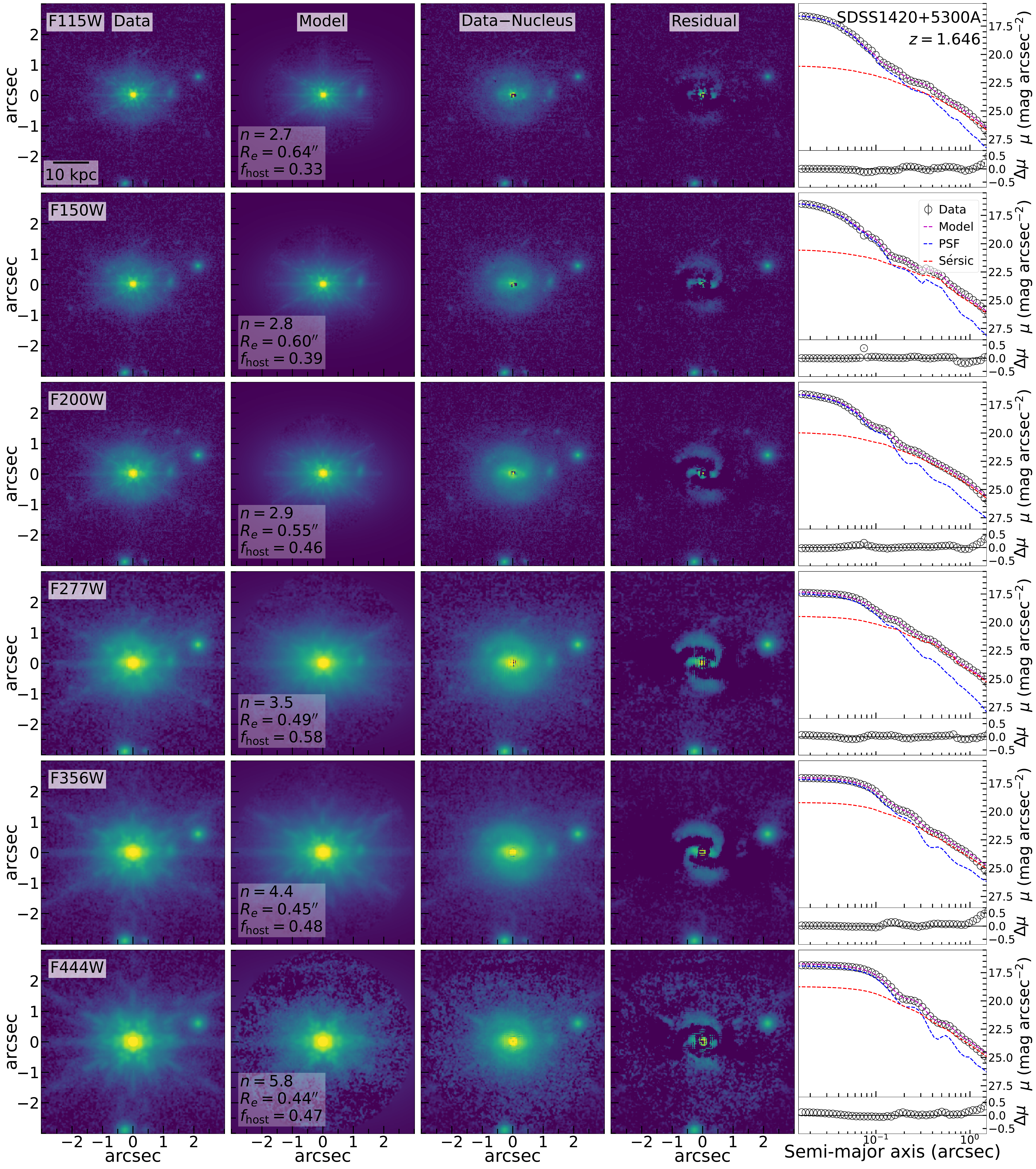}
\caption{Multiwavelength simultaneous AGN+host decomposition in six-band NIRCam images of the broad-line quasar SDSS1420+5300A at $z=1.646$ with the single \sersic\ component and the {\em global} PSF model (SS2) using \texttt{galfitm}. Images of data, model (AGN+host), data$-$nucleus, and residual (data$-$model) are shown from left to right. The best-fit parameters for the host galaxy and host-to-total flux fraction $f_{\rm host}$ are shown at the lower-left corner in the model panel. The right-most column shows the radial surface brightness ($\mu$) profiles of data and models, with the bottom panel showing the difference between data and model ($\Delta \mu=\mu_{\rm data}-\mu_{\rm model}$). The apparent spiral arm features in the residual panels indicate the presence of a stellar disk.}
\label{fig14}
\end{figure*}

\subsection{AGN-host Image Decomposition}\label{sec5.2}

We employ \texttt{galfitm} \citep{Haussler+2013MNRAS, Vika+2013MNRAS} to perform the AGN-host image decomposition. \texttt{galfitm} is a multiwavelength version of \texttt{galfit}, which can model images from multiple filters simultaneously, taking into account the wavelength-dependent galaxy structure. Works on nearby galaxies find that the \sersic\ index $n$ and $R_e$ tend to vary systematically with wavelength, with increasing $n$ and decreasing $R_e$ from rest-frame 0.3 to 2 \micron\ \citep[e.g.,][]{Kelvin+2012MNRAS}. This change of structure as a function of wavelength could result from variations in the spatial distributions of stellar populations, metallicity, and dust attenuation. 

We consider two sets of fitting configurations for the decomposition: (1) A PSF model for the AGN and a \sersic\ model for the host (single \sersic; SS); (2) A PSF model for the AGN, two \sersic\ models for the host (double \sersic; DS). For the DS fitting, the first \sersic\ model is for the elongated emission from the bar/lens-like component revealed in the residual image from fitting configuration 1, and the second \sersic\ model is for the emission from the disk-like component. Combining three options for PSF models, we have six sets of fits, denoted as SS1, SS2, SS3, DS1, DS2, DS3. We note that adding an additional bulge component is not evident. Our tests with a PSF+bulge+bar/lens+disk model suggests a bulge with $R_e\approx1$ pixel, whose emission is blended and dominated by the central point source.

We use a cutout size of 201 $\times$ 201 pixels ($\sim$6\arcsec), which is large enough to enclose most of the emission from our target. A close extended companion that is blended with our target is also fitted simultaneously with a \sersic\ model. All other sources in the cutout are masked. The centers of the AGN and the host are tied together and allowed to vary independently in different filters to account for residual astrometry mismatches ($\lesssim0.2$ pixel). For the SS fitting configuration, we use the same values for the ellipticity and position angle parameters across the wavelength, set \msersic\ and $m_{\rm AGN}$ free to vary (fifth-order polynomial), and allow the $n$ and $R_e$ to vary quadratically with wavelength, following \citet{Haussler+2013MNRAS}. $n$ is constrained to be in the range of 0.3 to 7. For the DS fitting configuration, we do not allow $n$ and $R_e$ to vary with wavelength for both \sersic\ models following \citet{Haussler+2022A&A}, with the assumption that no significant color gradient is present within each component. 

We set initial guesses of $R_e=10$ pixel, $n=3$, and \msersic\ and $m_{\rm AGN}$ to 1 mag fainter than the total magnitude for the SS configuration. For the DS configuration, we set $R_e=8$ pixel, $n=2$ for the first \sersic\ model, $R_e=20$ pixel, $n=1$ for the second \sersic\ model as initial guesses. The two \sersic\ models have the same initial \msersic, which is 0.75 mag larger than the best-fit \msersic\ from the SS fit. We perform the decomposition twice for both SS and DS configurations with the best-fit parameters of the first run as the initial guess for the second run. 

As mentioned in Section~\ref{sec4}, true uncertainties of the parameters are often larger than the formal uncertainties reported by the fitting code. We follow the procedures described above to generate mock AGNs using the best-fit parameters. For the sky background, we select a ``blank'' region in the mosaic near our target in each filter. The background cutout has the same size as the AGN and does not contain any visible source. We further remove residual sky pedestal to ensure that the $3\sigma$-clipped median is exactly zero. For each set, we generate 50 mock AGNs. The final uncertainty of each parameter is the quadrature sum of the uncertainty from the fit to observed data and the standard deviation from the fit to mock AGNs. 

Due to the bright magnitude of the host (19.2--21.3 mag, $\sim8$ mag brighter than the image depth) and moderate AGN-to-host ratio (0.7--2.6), the uncertainties in magnitudes are very small ($\lesssim0.01$ mag). There are also minor systematic biases between results from mock AGNs and input values. Systematic biases in host magnitude are $\sim$0.05 mag for SS2 and SS3 in the F444W filter, and $\lesssim0.06$ mag for DS1, DS2, and DS3 in all filters.  Systematic biases in $n$ are slightly larger, with $\sim9$\% for SS2 and SS3 in the F444W filter, 5\%, 5\%, and 0\% for $n$1 and 17\%, 13\%, and 14\% for $n$2 in DS1, DS2, and DS3, respectively. Almost no systematic bias ($\lesssim1$\%) is found for $R_e$ in any set of fits. The low systematic biases for parameters in SS1 are likely due to the large $n$, which hits the upper limit (7) in the F444W filter. The results from our six sets of multiwavelength simultaneous AGN+host image decomposition are summarized in Table~\ref{table3}. Systematic biases from the mock AGNs are not applied, as they are smaller than the differences due to adopting different PSF models or adopting different host models. For example, the differences of \msersic\ from using different PSF models are on the order of $\sim 0.1$~mag, while the differences from adopting different host models are on the order of $\sim 0.3$~mag. These latter systematic uncertainties are substantially larger than formal measurement uncertainties reported by the fit and scatter from mock AGNs. 

An example of decomposition with the single \sersic\ model and {\em global} PSF model (SS2) is presented in Figure~\ref{fig14}. Thanks to the excellent PSF of NIRCam, the total observed light at radius$\gtrsim$0\farcs2 in the SW filters is dominated by resolved emission from the host galaxy. We find prominent features of a central elongated structure, likely a bar or a lens, and galactic spiral arms in the data$-$nucleus images. The relatively high global \sersic\ index $n$ from three SS sets (3.2--4.0) in the F277W filter, which corresponds to rest-frame $\sim 1$\micron, is primarily due to the combination of a compact high-surface-brightness bar/lens and an extended low-surface-brightness disk. On the other hand, $n$ of the second \sersic\ component in the three DS sets is $\sim 1.0$, consistent with the value for exponential disks ($n=1$). Our results highlight the danger of morphological classifications based solely on $n$ from single \sersic\ model fits with the presence of resolved substructures in the host galaxy, in particular high surface brightness compact structures such as a bar/lens. 

\begin{deluxetable*}{cCCCCCC}
\caption{AGN-host Image Decomposition Results of SDSS1420+5300A \label{table3}}
\tabletypesize{\scriptsize}
\tablehead{
\colhead{Parameter} & \colhead{SS1} & \colhead{SS2} & \colhead{SS3}& \colhead{DS1} & \colhead{DS2} & \colhead{DS3}\\
\colhead{(1)} & \colhead{(2)} & \colhead{(3)} & \colhead{(4)} & \colhead{(5)} & \colhead{(6)} & \colhead{(7)}
}
\startdata
\multicolumn{7}{c}{F115W}\\
AGN-to-host ratio &  1.87            &  2.05            &  2.17             &  2.26           &  2.21           &  2.59          \\
$m_{\rm AGN}$ (mag)& 20.56   \pm 0.01 & 20.52   \pm 0.01 & 20.51    \pm 0.01 & 20.55 \pm  0.01 & 20.54 \pm  0.01 &  20.50 \pm  0.01\\ 
\msersic1 (mag) & 21.24   \pm 0.01 &  21.30   \pm 0.01 & 21.35    \pm 0.01 & 22.62 \pm  0.01 & 22.42 \pm  0.01 & 22.99 \pm  0.01\\ 
$R_e$1 (\arcsec) & 0.618     \pm    0.008 & 0.644     \pm    0.008 & 0.648      \pm    0.007 &   0.207 \pm     0.001 &   0.189 \pm     0.001 &   0.252 \pm     0.001\\
$n$1 &  3.16 \pm  0.05 &  2.74 \pm  0.06 &  2.52 \pm  0.06  & 1.16 \pm  0.01 &  2.13 \pm  0.02 &  0.58  \pm 0.01\\
\msersic2 (mag) & \nodata          & \nodata          &  \nodata          & 21.88 \pm  0.01 & 21.94 \pm  0.01 & 21.86 \pm  0.01\\
$R_e$2 (\arcsec) & \nodata          & \nodata          &  \nodata          &   0.650 \pm     0.001 &   0.644 \pm     0.003 &   0.657 \pm     0.001\\
$n$2 & \nodata          & \nodata          &  \nodata          &  0.89 \pm  0.01 &  0.96 \pm  0.01 &  0.88 \pm  0.01\\
\hline
\multicolumn{7}{c}{F150W}\\
AGN-to-host ratio &  1.49            &  1.58            &  1.64             &   1.8           &  1.67           &  1.93          \\ 
$m_{\rm AGN}$ (mag)& 20.43   \pm 0.01 & 20.42   \pm 0.01 & 20.41    \pm 0.01 & 20.41 \pm  0.01 & 20.44 \pm  0.01 & 20.39 \pm  0.01\\ 
\msersic1 (mag) & 20.86   \pm 0.01 & 20.92   \pm 0.01 & 20.95    \pm 0.01 & 22.42 \pm  0.02 & 22.13 \pm  0.01 & 22.59 \pm  0.01\\ 
$R_e$1 (\arcsec) & 0.592     \pm    0.005 & 0.601     \pm    0.006 & 0.602      \pm    0.005 &   0.207 \pm     0.001 &   0.189 \pm     0.001 &   0.252 \pm     0.001\\
$n$1 &  3.17 \pm  0.03 &  2.76 \pm  0.04 &  2.53 \pm  0.04  & 1.16 \pm  0.01 &  2.13 \pm  0.02 &  0.58  \pm 0.01\\
\msersic2 (mag) & \nodata          & \nodata          &  \nodata          & 21.41 \pm  0.01 & 21.47 \pm  0.01 & 21.42 \pm  0.01\\
$R_e$2 (\arcsec) & \nodata          & \nodata          &  \nodata          &   0.650 \pm     0.001 &   0.644 \pm     0.003 &   0.657 \pm     0.001\\
$n$2 & \nodata          & \nodata          &  \nodata          &  0.89 \pm  0.01 &  0.96 \pm  0.01 &  0.88 \pm  0.01\\
\hline
\multicolumn{7}{c}{F200W}\\
AGN-to-host ratio &  1.05            &  1.17            &  1.22             &  1.27           &  1.22           &  1.45          \\
$m_{\rm AGN}$ (mag)& 20.32   \pm 0.01 & 20.27   \pm 0.01 & 20.26    \pm 0.01 &  20.30 \pm  0.01 &  20.30 \pm  0.01 & 20.23 \pm  0.01\\ 
\msersic1 (mag) & 20.37   \pm 0.01 & 20.44   \pm 0.01 & 20.48    \pm 0.01 & 21.65 \pm  0.01 & 21.48 \pm  0.01 & 21.93 \pm  0.01\\ 
$R_e$1 (\arcsec) & 0.558     \pm    0.003 & 0.549     \pm    0.003 & 0.548      \pm    0.003 &   0.207 \pm     0.001 &   0.189 \pm     0.001 &   0.252 \pm     0.001\\
$n$1 &  3.35 \pm  0.03 &  2.91 \pm  0.03 &  2.65 \pm  0.03  & 1.16 \pm  0.01 &  2.13 \pm  0.02 &  0.58  \pm 0.01\\
\msersic2 (mag) & \nodata          & \nodata          &  \nodata          & 21.05 \pm  0.01 & 21.09 \pm  0.01 & 21.02 \pm  0.01\\
$R_e$2 (\arcsec) & \nodata          & \nodata          &  \nodata          &   0.650 \pm     0.001 &   0.644 \pm     0.003 &   0.657 \pm     0.001\\
$n$2 & \nodata          & \nodata          &  \nodata          &  0.89 \pm  0.01 &  0.96 \pm  0.01 &  0.88 \pm  0.01\\
\hline
\multicolumn{7}{c}{F277W}\\
AGN-to-host ratio &  0.68            &  0.71            &  0.77             &  0.90           &  0.73           &  1.00          \\
$m_{\rm AGN}$ (mag)& 20.26  \pm 0.01 & 20.26 \pm  0.01 & 20.22 \pm  0.01 & 20.19 \pm  0.01 & 20.31 \pm  0.01 & 20.13  \pm 0.01\\
\msersic1 (mag) & 19.84  \pm 0.01 & 19.89 \pm  0.01 & 19.94 \pm  0.01 & 20.98 \pm  0.01 &  20.70 \pm  0.01 & 21.22 \pm  0.01\\
$R_e$1 (\arcsec) & 0.508     \pm    0.002 & 0.487     \pm    0.002 & 0.489      \pm    0.002 &   0.207 \pm     0.001 &   0.189 \pm     0.001 &   0.252 \pm     0.001\\
$n$1 &  4.04 \pm  0.02 &  3.46 \pm  0.03 &  3.15 \pm  0.03  & 1.16 \pm  0.01 &  2.13 \pm  0.02 &  0.58  \pm 0.01\\
\msersic2 (mag) & \nodata          & \nodata          &  \nodata          & 20.70 \pm  0.01 & 20.74 \pm  0.01 & 20.63 \pm  0.01\\
$R_e$2 (\arcsec) & \nodata          & \nodata          &  \nodata          &   0.650 \pm     0.001 &   0.644 \pm     0.003 &   0.657 \pm     0.001\\
$n$2 & \nodata          & \nodata          &  \nodata          &  0.89 \pm  0.01 &  0.96 \pm  0.01 &  0.88 \pm  0.01\\
\hline
\multicolumn{7}{c}{F356W}\\
AGN-to-host ratio &   1.0            &  1.08            &  1.15             &  1.46           &  1.25           &   1.6          \\
$m_{\rm AGN}$ (mag)& 19.61   \pm 0.01 & 19.58   \pm 0.01 & 19.55    \pm 0.01 & 19.52 \pm  0.01 & 19.57 \pm  0.01 & 19.46 \pm  0.01\\ 
\msersic1 (mag) & 19.61   \pm 0.01 & 19.66   \pm 0.01 &  19.70    \pm 0.01 & 20.88 \pm  0.01 & 20.53 \pm  0.01 & 21.02 \pm  0.01\\ 
$R_e$1 (\arcsec) & 0.467     \pm    0.003 & 0.452     \pm    0.002 & 0.464      \pm    0.002 &   0.207 \pm     0.001 &   0.189 \pm     0.001 &   0.252 \pm     0.001\\
$n$1 &  5.19 \pm  0.04 &  4.37 \pm  0.04 &  3.99 \pm  0.04  & 1.16 \pm  0.01 &  2.13 \pm  0.02 &  0.58  \pm 0.01\\
\msersic2 (mag) & \nodata          & \nodata          &  \nodata          & 20.52 \pm  0.01 &  20.60 \pm  0.01 & 20.49 \pm  0.01\\
$R_e$2 (\arcsec) & \nodata          & \nodata          &  \nodata          &   0.650 \pm     0.001 &   0.644 \pm     0.003 &   0.657 \pm     0.001\\
$n$2 & \nodata          & \nodata          &  \nodata          &  0.89 \pm  0.01 &  0.96 \pm  0.01 &  0.88 \pm  0.01\\
\hline
\multicolumn{7}{c}{F444W}\\
AGN-to-host ratio &  1.02            &  1.12            &  1.21             &  1.78           &  1.55           &   2.0          \\
$m_{\rm AGN}$ (mag)& 19.18   \pm 0.01 & 19.15   \pm 0.01 & 19.11    \pm 0.01 & 19.04 \pm  0.01 & 19.09 \pm  0.01 & 18.99 \pm  0.01\\ 
\msersic1 (mag) &  19.20   \pm 0.01 & 19.27   \pm 0.01 & 19.32    \pm 0.01 & 20.49 \pm  0.01 & 20.25 \pm  0.01 & 20.79 \pm  0.01\\ 
$R_e$1 (\arcsec) &   0.430   \pm   0.005  &   0.444   \pm    0.005 &   0.475    \pm    0.005 &   0.207 \pm     0.001 &   0.189 \pm     0.001 &   0.252 \pm     0.001\\
$n$1 &   7.0 \pm  0.09 &   5.8 \pm  0.11 &  5.33 \pm   0.10  & 1.16 \pm  0.01 &  2.13 \pm  0.02 &  0.58  \pm 0.01\\
\msersic2 (mag) & \nodata          & \nodata          &  \nodata          & 20.35 \pm  0.01 & 20.39 \pm  0.01 & 20.26 \pm  0.01\\
$R_e$2 (\arcsec) & \nodata          & \nodata          &  \nodata          & 0.650   \pm  0.001    & 0.644   \pm    0.003  &  0.657  \pm     0.001\\
$n$2 & \nodata          & \nodata          &  \nodata          &  0.89 \pm  0.01 &  0.96 \pm  0.01 &  0.88 \pm  0.01\\
\hline
\multicolumn{7}{c}{Goodness of Image Decomposition}\\
$\chi^2_{\nu}$ & 5.50 & 5.87 & 5.55 & 4.83 & 5.18 & 4.65\\
\hline
\multicolumn{7}{c}{Stellar Mass}\\
$\log M_*$ ($M_{\odot}$) & $11.25\pm0.07$ & $11.23\pm0.07$ & $11.21\pm0.07$ & $11.09\pm0.09$ & $11.15\pm0.08$& $11.08\pm0.09$\\
\enddata
\tablecomments{Derived parameters from our multiwavelength simultaneous AGN-host image decomposition in each filter. Results from two configurations (single \sersic\ and double \sersic) with three PSF models are shown in the Columns (2-7), respectively. Uncertainty is the quadrature sum of the formal uncertainty from fit to real AGN images and $1\sigma$ standard deviation of best-fit results to mock AGNs. A minimum uncertainty of 0.01 is adopted for magnitude and $n$, and 0.001 for $R_e$. These uncertainties are much smaller than the systematic biases from PSF and model mismatch. Stellar masses derived from SED fitting to AGN contamination-subtracted host fluxes in six NIRCam filters are shown in the bottom row. }
\end{deluxetable*}

\begin{figure}[t]
\centering
\includegraphics[width=0.5\textwidth]{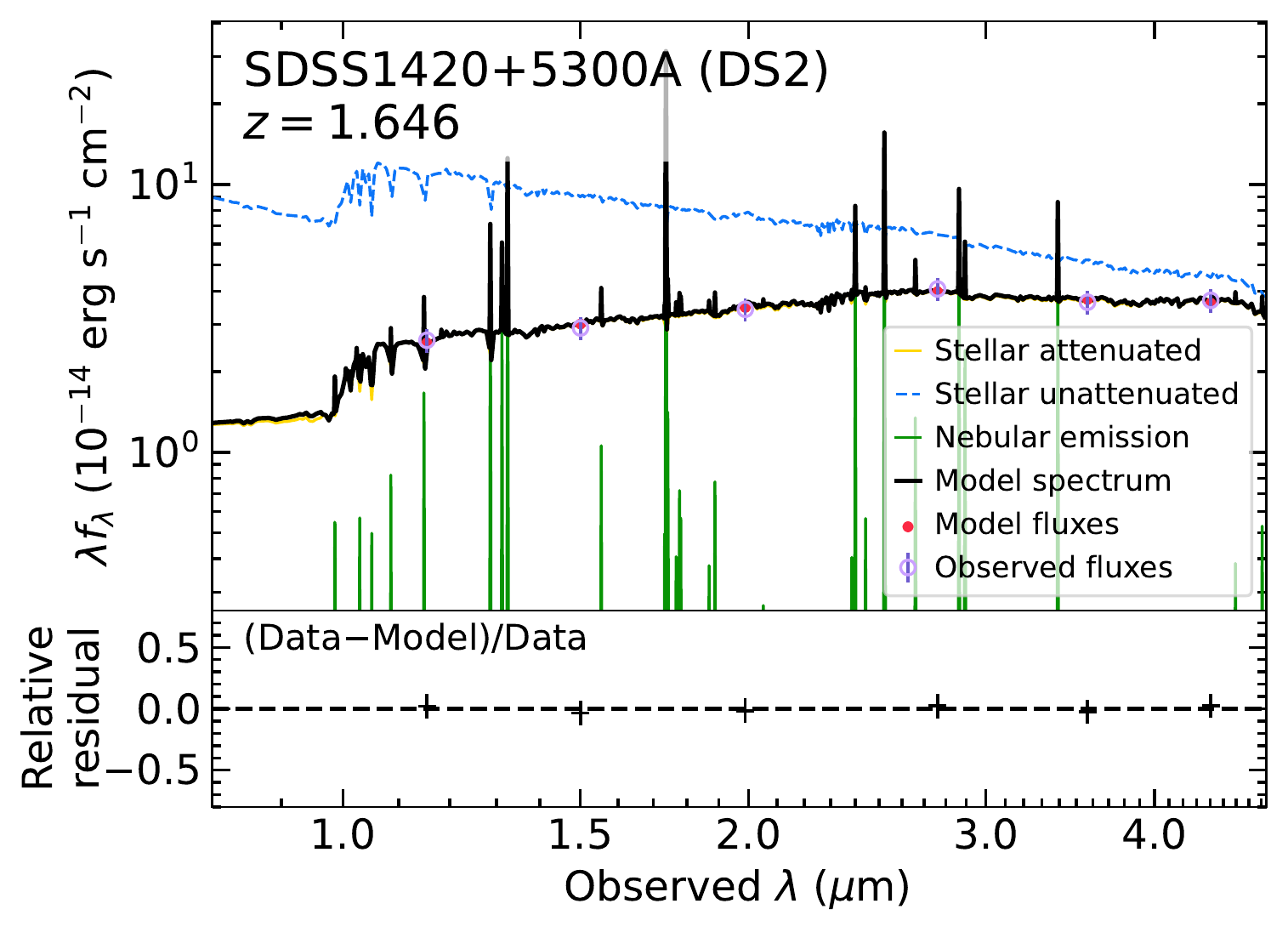}
\caption{{SED fitting to the AGN-subtracted host photometry in the six NRICam filters from decomposition configuration DS2 using \texttt{CIGALE}. The observed fluxes are shown in purple open circles. The best-fit total model spectrum is shown in the black solid curve, with its components of stellar attenuated emission and nebular emission shown in the yellow and green solid curves. Red filled circles represent the model fluxes and the dashed blue curve represents the stellar emission before attenuation. Relative residuals (Data$-$Model)/Data are shown in the lower panel.}}
\label{fig15}
\end{figure}

\subsection{Stellar Mass Estimation}\label{sec5.3}

We correct the foreground galactic extinction adopting the extinction curve from \citet{CCM1989} with $R_V=3.1$ and the extinction map from \citet{Schlegel+1998ApJ}. As mentioned in \cite{Bagley+2022CEERS_data_reduction}, the relative and absolute photometric calibration accuracy of the NIRCam images in their data product were still at the level of $\sim5$\%. We assume a conservative 10\% uncertainty for host fluxes. We then derive the stellar mass by feeding AGN contamination-subtracted host galaxy fluxes to the spectral energy distribution (SED) fitting code \texttt{CIGALE v2022.1} \citep{Boquien+2019A&A, Yang+2022ApJ}. {The SED fitting is performed on all decomposition model sets (Table~\ref{table3}).}

For the models of \texttt{CIGALE}, we adopt a single stellar population model from \citet{Bruzual&Charlot2003MNRAS} with solar metallicity, a \citet{Chabrier2003PASP} IMF, and a ``delayed'' star formation history. We also include a \texttt{nebular} emission line model with ionization parameter $U=10^{-2}$, and a \texttt{dustatt\_modified\_starburst} model with extinction curve adapted from \citet{Calzetti+2000ApJ}. We adopt a wide range of stellar ages: 0.50--3.85 Gyr with a step of 0.05 Gyr; an e-folding time of the main stellar population (Gyr): 0.001, 0.05, 0.1, 0.25, 0.5, 0.75, 1.0, 1.25, 1.5, 2.0, 2.5, 3.0, 3.5, 4.0, 4.5, 5.0, 6.0, 7.0, 8.0, 9.0, 10, 11, 12, 13, 14, 15, 16, 17, 18, 19, 20; and color excess of the nebular lines [$E(B-V)$ (mag)]: 0, 0.001, 0.005, 0.01, 0.03, 0.05, 0.1, 0.2, 0.3, 0.4, 0.5, 0.6, 0.7. All other parameters are fixed with their default values. Since the longest filter F444W only covers up to rest-frame $1.7$~\micron, we do not consider dust emission models. An example of SED fitting to host photometries from DS2 is presented in Figure~\ref{fig15}.

We adopt a Bayesian estimate of the stellar mass, which is based on the probability density distribution of the likelihood of all templates. Uncertainties of the stellar masses are 0.08-0.09 dex ($\sim18-21$\%) for all six sets of fits. This level of uncertainty is larger than that of the input flux (10\%), suggesting that the uncertainty is dominated by the degeneracy of the templates. {However, we caution that the systematic uncertainties due to the assumptions in the fitting, such as the IMF and star formation history, can be significantly larger than the formal stellar mass uncertainties reported by the fitting \citep[e.g.,][]{Conroy2013ARA&A, Lower+2020ApJ}.} Stellar mass measurements are presented in Table~\ref{table3}.

\begin{figure}[t]
\centering
\includegraphics[width=0.5\textwidth]{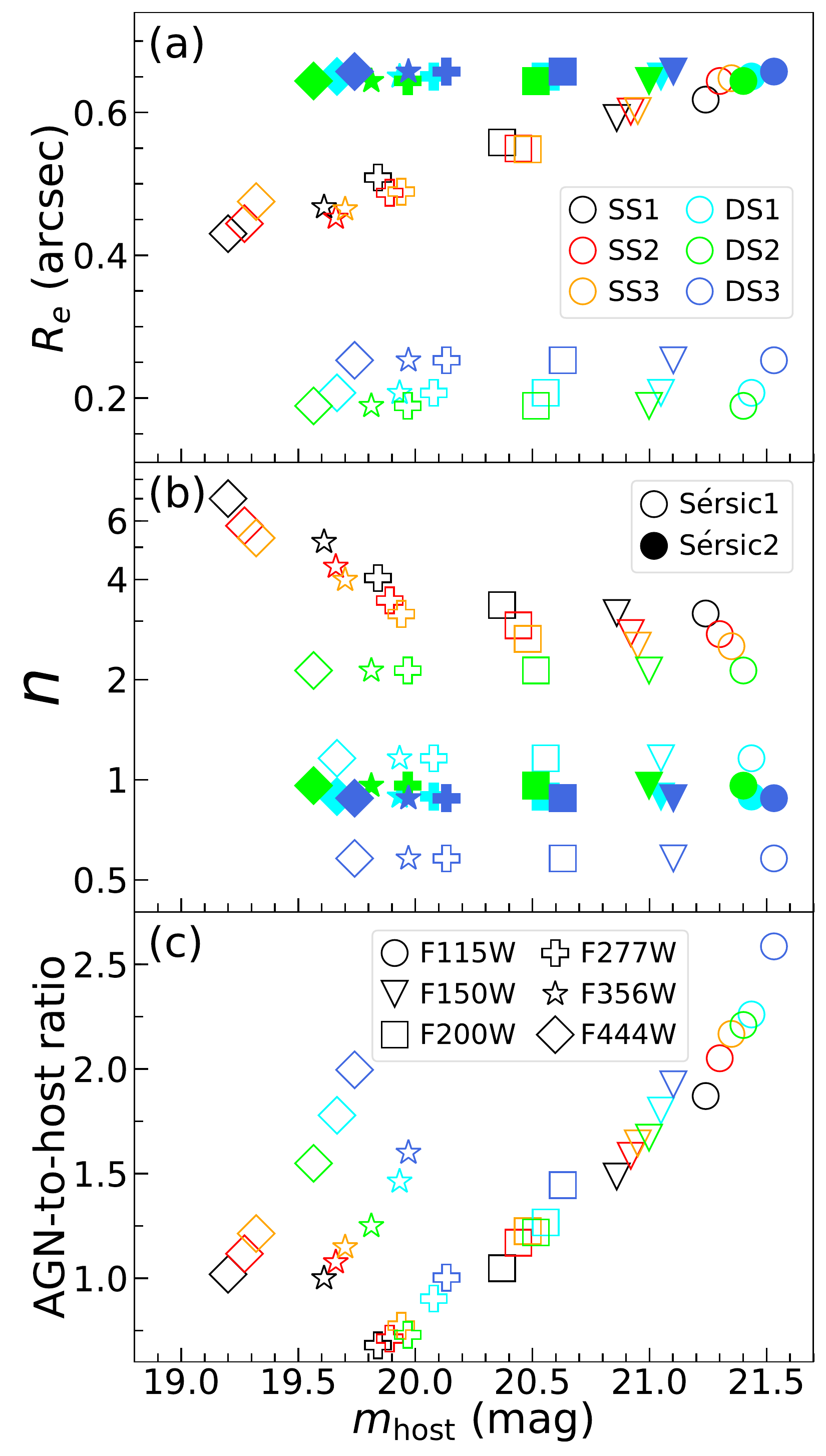}
\caption{Comparison of best-fit parameters (a) $R_e$, (b) $n$, and (c) AGN-to-host ratio versus total host magnitude ($m_{\rm host}$) from six sets of decomposition configurations. Black, red, orange, cyan, green, and blue symbols represent results from SS1, SS2, SS3, DS1, DS2, and DS3, respectively. Open symbols represent results of \sersic1, and filled symbols represent results of \sersic2. Circle, triangle, square, cross, star, and diamond symbols represent results in F115W, F150W, F200W, F277W, F356W, and F444W, respectively. Individual uncertainties are smaller than the size of symbols and thus are not shown.}
\label{fig16}
\end{figure}

\subsection{The Effect of PSF and Model Mismatch}

The bright emission, large size and relatively high contribution of the host galaxy in SDSS1420$+$5300A enable us to investigate the effects of PSF mismatch and host model mismatch on the derived host properties. 

As mentioned in Section~\ref{sec5.1}, the FWHM of PSF models is generally in ascending order of SS1-SS2-SS3 (DS1-DS2-DS3). For the three SS sets, we find clear trends of increasing host flux, decreasing AGN flux, and increasing $n$ toward narrower PSF models. The difference among the three SS sets is $\sim0.1$ mag in \msersic, 0.02--0.07 mag in $m_{\rm AGN}$, 9--16\% in AGN-to-host ratio, 1--9\% in $R_e$, and $\sim25$\% for $n$. The overall trends of these parameters are consistent with the mock analysis presented in Section~\ref{sec4}. Due to the complication of two \sersic\ models, the trends of parameters in the DS sets are not as clear as for the SS sets. The parameter differences are larger for the first \sersic\ model, likely a result from the larger AGN-to-host ratio and smaller host size. A comparison of the best-fit parameters from six sets of decomposition configurations is shown in Figure~\ref{fig16}.

Since host galaxies are fitted with two \sersic\ models in the DS sets, we could only compare the total AGN magnitude, total host magnitude and stellar mass between SS and DS sets. We find that host magnitudes tend to be larger (lower flux) while AGN magnitudes tend to be smaller (higher flux) in DS models than in SS models. These differences lead to $\sim15-60$\% higher AGN-to-host ratio from F115W to F444W in DS sets compared to SS sets. The stellar masses are $\sim0.11$ dex (29\%) lower in the DS sets compared to the SS sets. This offset is larger than the uncertainty of individual values (18--21\%) and the differences among the three PSF models for the same model configuration (0.04 dex for SS and 0.07 dex for DS). $\chi^2_{\nu}$ is smaller (better fits) for DS sets. The improvement is expected given the addition of an extra component and more free parameters. Fits from all three PSF models have similar $\chi^2_{\nu}$, with broadened {\em global} PSF models performing slightly better. Due to the overall similar $\chi^2_{\nu}$, we cannot determine which PSF model gives the best results. However, it is likely that the broadened {\em global} PSF model provides less biased results according to the mock analysis in Section~\ref{sec4.2}. 

For stellar mass, we find that the effect of model mismatch (0.11 dex on average) is larger than the effect of PSF mismatch (0.04 dex for SS and 0.07 dex for DS). We would expect more significant impact of PSF mismatch in systems with more AGN-dominated light. 

\subsection{Comparison with Earlier Work}

The NIRCam imaging data of SDSS1420+5300A were also analyzed by \citet{Ding+2022ApJ} using \texttt{galight} with a single \sersic\ model for the host emission. There are a few key differences in the PSF construction and image decomposition settings. \citet{Ding+2022ApJ} construct a PSF library by sorting FWHM of bright point sources and selecting isolated and sharp ones in the FoV of four pointings. They then use the top 3, 5, and 8 PSF models with good performance in terms of $\chi^2_{\nu}$ of the fit and construct a combined average PSF model by stacking them. The final reported measurements are performed using either individual PSF models or combined PSF models based on the smallest $\chi^2_{\nu}$ value, with uncertainties derived from the dispersion of results using the top-5 PSF models. They perform image decomposition independently for each filter, so that they can use a finer pixel scale of 0\farcs015 in SW fitlers. 

Since the photometric calibration reference file in \citet{Ding+2022ApJ} is different with the one from \citet{Bagley+2022CEERS_data_reduction}, we only compare their AGN-to-host ratio, $R_e$, and $n$ with those from our SS configurations. Their $R_e$ is $\sim20\%$ smaller than ours in four NIRCam filters listed in their paper (F150W, F200W, F356W, and F444W). Their $n$ is approximately constant around $2$ in all four filters. This is inconsistent with the $n$ values from our SS sets, which show a clear positive wavelength dependence (e.g., 2.7 to 5.8 for SS2), as expected since older stellar populations have more concentrated distributions. We also find that their AGN-to-host ratios are systematic higher than ours in all four filters. Despite the fainter host magnitudes, their derived stellar mass is on average $\sim0.12$ and 0.25 dex higher than that from our SS and DS sets, respectively. Since we adopt the same cosmological parameter and IMF, it is likely that at least part of the systematic difference is due to the different SED fitting codes used. 

\begin{figure}[t]
\centering
\includegraphics[width=0.4\textwidth]{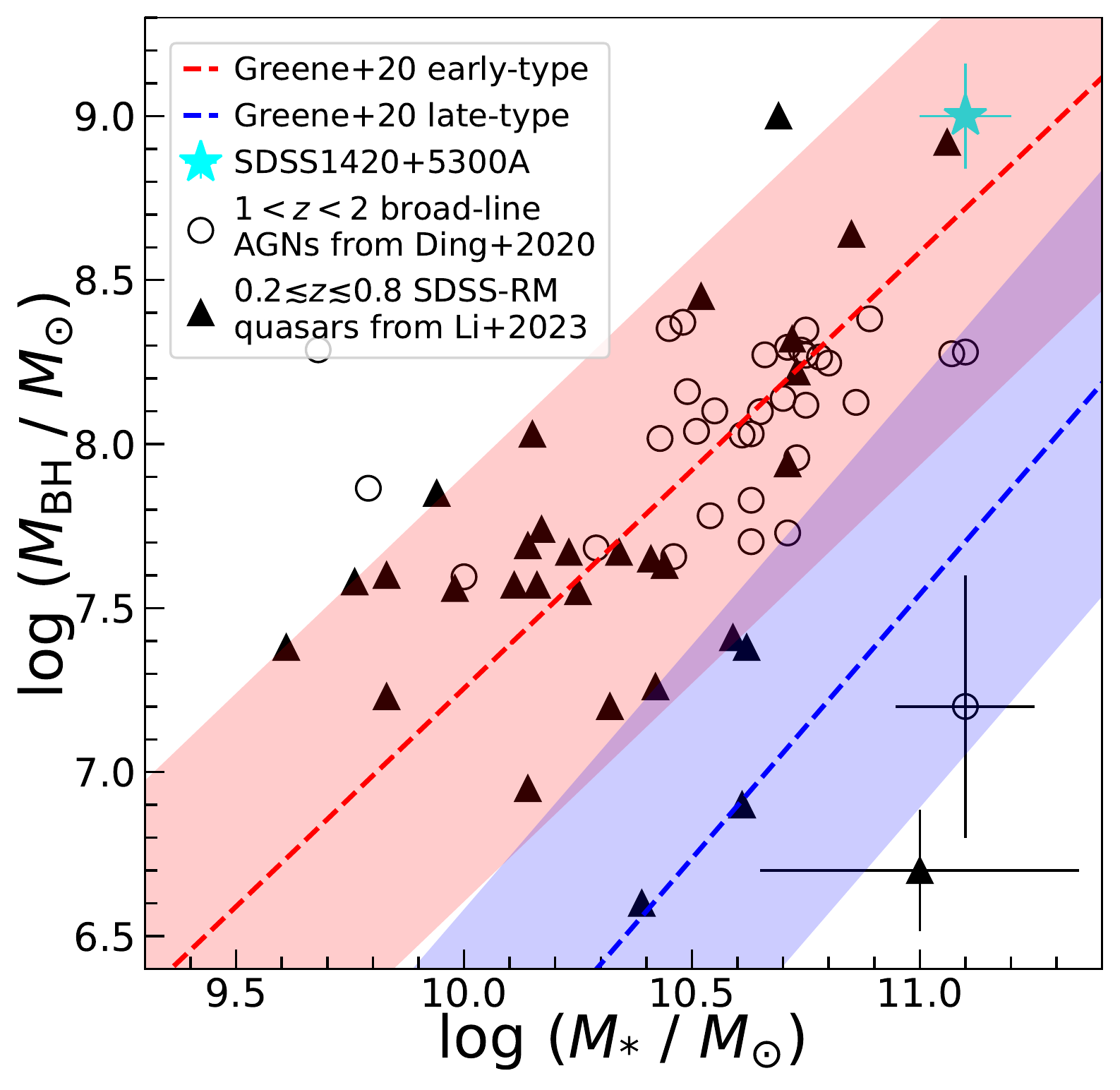}
\caption{Comparison of $M_{\rm BH}$ versus $M_*$ relation of SDSS1420+5300A (cyan star) with local relations from inactive galaxies \citep{Greene+2020ARA&A}. Red and blue dashed lines represent local relations derived from early-type and late-type galaxies, respectively, with shaded region indicating the intrinsic scatter. Open circles represent X-ray selected $1<z<2$ broad-line AGNs from \cite{Ding+2020ApJ}. We recalculate their $M_{\rm BH}$ using Equation 5 from \citet{Woo+2015ApJ} to match the choice of virial factor adopted for SDSS1420+5300A in \cite{Grier+2019ApJ}. Filled triangles represent SDSS-RM quasars at $0.2\lesssim z \lesssim0.8$ with $M_{\rm BH}$ from reverberation mapping in \citet{Li+2023arXiv}. Typical uncertainties are indicated at the lower-right corner.}
\label{fig17}
\end{figure}

\subsection{SDSS1420+5300A on the $M_{\rm BH}-M_*$ Plane}

Previous work on the $M_{\rm BH}-M_*$ relation of $z\approx2$ AGNs primarily relied on HST imaging for AGN-host decomposition \citep[e.g.,][]{Jahnke+2009ApJ, Bennert+2011ApJ, Schramm&Silverman2013ApJ, Ding+2020ApJ}, with a few studies relying on SED decomposition \citep[e.g.,][]{Bongiorno+2014MNRAS, Suh+2020ApJ}. The lack of rest-frame NIR coverage ($>0.5$\micron) and $\sim1.6$ kpc resolution of HST make the decomposition and stellar mass measurements less robust. Also, the $M_{\rm BH}$ estimates in previous studies are mostly based on the single-epoch virial method, which may suffer from complicated systematics resulting from factors such as the poorly understood size-luminosity of the broad-line region at high-$z$, statistical biases due to sample selection, and different lines adopted in the BH mass estimation (e.g., \MgII\ and \CIV) other than \Hb\ \citep[e.g.,][]{Shen_2013, Shen+2015ApJS}. 

With a more robust $M_{\rm BH}$ estimate from the Sloan Digital Sky Survey Reverberation Mapping project \citep{Shen+2015ApJS, Grier+2019ApJ}, we can now place SDSS1420+5300A on the $M_{\rm BH}-M_*$ plane. We adopt the average $M_*$ estimate from the DS sets, $M_*=10^{11.1 \pm 0.1} M_{\odot}$, given their better modeling of the host structure. Comparing with the local $M_{\rm BH}-M_*$ relation from inactive galaxies in \citet{Greene+2020ARA&A}, we find that this object lies $\sim0.6$ dex above the upper envelope (best-fit relation + $1\times$ intrinsic scatter) of local late-type (spiral) galaxies, but consistent with early-type (elliptical and S0) galaxies (Figure~\ref{fig17}). {The position of this object is in broad agreement with previous studies with host fluxes derived from image decomposition using X-ray selected broad-line AGNs at similar redshifts \citep[e.g.,][]{Jahnke+2009ApJ, Bennert+2011ApJ, Schramm&Silverman2013ApJ, Ding+2020ApJ} and SDSS-RM quasars at $0.2\lesssim z \lesssim0.8$ \citep{Li+2023arXiv} with $M_{\rm BH}$ from reverberation mapping}. An increase of at least 0.4 dex in $M_*$ is required to reconcile this object with the local relation for late-type galaxies. This amount of growth is feasible as local Sab-Scd galaxies are found to double their mass within 8 Gyr \citep{Bellstedt+2020MNRAS}. Alternatively, gas-rich major merger, which is more common at high redshift \citep[e.g.,][]{Lotz+2011ApJ}, could change the morphology of the object and bring it back to the local relation. A more comprehensive analysis of the $M_{\rm BH}-M_*$ relation at $z\approx2$ requires a larger sample with well understood selection functions, reliable BH masses from reverberation mapping and stellar masses from JWST imaging.

\section{Summary}\label{sec6}

In this work, we characterize the properties of the NIRCam PSF and their spatial variations in eight filters (F070W, F115W, F150W, F200W, F277W, F356W, F444W, and F480M), using public NIRCam imaging data. Using simulated AGN+host images, we investigate the recoverability of host properties with imaging decomposition, and the impact of mismatched PSF models. As an illustration for our methodology, we applied the optimal PSF construction method to NIRCam images in the CEERS field, and measured host properties for the broad-line AGN SDSS1420+5300A at $z=1.646$. Our main results are as follows: 

\begin{itemize}

\item{By comparing three commonly-used methods for PSF construction, we find that PSF models constructed with \texttt{PSFEx} have the the best quality. \texttt{PSFEx} surpasses the other two methods by producing PSF models with the highest SNR and the best performance in modeling the light profiles of point sources in the image.}

\item{We find clear PSF spatial variations across the FoV of NIRCam, with maximum and RMS variations of the FWHM decreasing from $\sim20$\% and 5\% in the F070W filter to $\sim3$\% and 0.6\% in the F444W filter. These spatial variations of NIRCam PSFs can have significant consequences on the imaging decomposition of broad-line AGN+host systems.}

\item{We recommend the use of \texttt{PSFEx} to construct the PSF model, as well as modeling its spatial variations. If a sufficient number of point sources are available, it is recommended to construct {\em region} or {\em local} PSF models. Otherwise, adopting the {\em global} PSF model still provides satisfactory results under most circumstances, with an average $\lesssim0.02$ mag systematic offset and $\lesssim0.05$ mag random scatter. }


\item{We generate simulated images of mock AGNs spanning a wide range of parameters to investigate the reliability and biases of parameter recovery from AGN-host image decomposition using perfect, narrower and broader PSF models. The narrower or broader PSF models are representative of the maximum spatial variations (e.g., 3--20\% fractional changes in the FWHM) of the PSF found in this study. PSF mismatch has a profound impact on the recovered host properties, leading to larger scatter, greater systematic biases, and fewer objects with successful measurements compared with the results from using a perfect PSF model (which is unavailable in realistic situations). 

Fluxes of host galaxies are more likely overestimated using either a broader or a narrower PSF model. The broader PSF model tends to produce flatter host galaxies (smaller $n$), while the narrower PSF model tends to produce more concentrated and smaller host galaxies (larger $n$ and smaller $R_e$). Narrower PSF models on average have a much worse impact on the recovery of host properties than broader PSF models. The systematic biases in the recovered host parameters generally increase as the AGN-to-host ratio increases. In cases where the AGN light dominates, the host flux can be overestimated by as much as a factor of few using PSF model narrower than the actual PSF at the location of the target. }

\item{Given the higher measurement success rate and lower biases for recovered host parameters with broader PSF models than with narrower PSF models, a practical suggestion is to slightly broaden the PSF model by the typical level of spatial variations across the FoV when very few point sources are available for accurate modeling of the PSF spatial variation. This strategy can mitigate the adverse effects and reduce significant biases from inadvertently adopting a narrower PSF model in the AGN+host decomposition.}
 
\item{The deviations of recovered parameters from the input values are larger than the formal measurement uncertainties for the vast majority of cases where
successful measurements are obtained (measurement/uncertainty$>3$). For example, median $|$deviation$|$/uncertainty of \msersic\ increases from $\sim1.2$ for AGN-to-host ratio=0.1 to 5.0 for AGN-to-host ratio=10, with extreme values as high as $\sim100$. It is strongly recommended to generate object-tailored mock AGNs with input parameters from the best fit to evaluate systematic biases and to obtain more realistic measurement uncertainties.}

\item{When the centers of the AGN and its host galaxy are not tied together, one can easily measure an artificial offset between them, dominated by the offset of the fitted host center. This artificial offset scales with the host surface brightness and is approximately aligned with the position angle of the host galaxy. The artificial offsets can be as large as $\sim80$\%, 26\%, and 7\% of $R_e$ in objects with $R_e=$0\farcs12, 0\farcs48, and 1\farcs92 when approaching the SB detection limit, respectively.}

\item{We find that the effects of PSF mismatch on the recovered properties of real AGN hosts (e.g., SDSS1420$+$5300A) are similar to those found in the mock data. The presence of small-scale high-surface-brightness substructures (e.g., bar and lens) would lead to high \sersic\ indices $n$ from the PSF+single \sersic\ fitting. Morphological classifications solely based on $n$ can lead to incorrect results. The impact of model mismatch (i.e., using incorrect model components for the host galaxy) may be greater than that of PSF mismatch in systems where the light contribution from the host galaxy is comparable to or larger than that from the AGN, particularly in the presence of substructures.}


\end{itemize}

\begin{acknowledgments}
We thank the referee for the constructive comments and Fengwu Sun for help on astrometry correction. This work is partially supported by NSF grants AST-2009947 and AST-2108162. Based on observations with the NASA/ESA/CSA James Webb Space Telescope obtained from the MAST archive at the Space Telescope Science Institute, which is operated by the Association of Universities for Research in Astronomy, Incorporated, under NASA contract NAS5- 03127. Support for Program number JWST-GO-02057 was provided through a grant from the STScI under NASA contract NAS5- 03127.
\end{acknowledgments}

%

\vspace{5mm}


\software{\texttt{Astropy} \citep{2013A&A...558A..33A,2018AJ....156..123A},  
          \texttt{GALFIT} \citep{Peng+2002AJ, Peng+2010AJ},
          \texttt{GALFITM} \citep{Haussler+2013MNRAS, Vika+2013MNRAS},
          \texttt{Matplotlib} \citep{Hunter2007}, 
          \texttt{Numpy} \citep{Harris2020}, 
          \texttt{photutils} \citep{photutils},
          \texttt{PSFEx} \citep{Bertin2011ASPC}, 
          \texttt{scipy} \citep{scipy}, 
          \texttt{SExtractor} \citep{1996A&AS..117..393B}, 
          \texttt{SWarp} \citep{Bertin+2002ASPC}}




\bibliography{sample631}{}
\bibliographystyle{aasjournal}

\end{CJK*}
\end{document}